\pdfoutput=1
\documentclass[aps,pra,reprint,superscriptaddress,nofootinbib,floatfix,showkeys]{revtex4-2}

\usepackage[T1]{fontenc}
\usepackage{amsmath,amssymb,amsthm,mathtools,bm}
\usepackage{booktabs,graphicx,xcolor,hyperref,orcidlink}
\usepackage{tikz}
\usetikzlibrary{arrows.meta,positioning,calc}
\usetikzlibrary{arrows.meta,positioning}
\hypersetup{colorlinks=true,linkcolor=blue,citecolor=blue,urlcolor=blue}
\allowdisplaybreaks
\graphicspath{{figures/}{./}}
\usepackage{placeins}

\newcommand{\ii}{\mathrm{i}}
\newcommand{\dd}{\mathrm{d}}
\newcommand{\ee}{\mathrm{e}}
\newcommand{\Tr}{\operatorname{Tr}}
\newcommand{\Var}{\operatorname{Var}}
\newcommand{\diag}{\operatorname{diag}}
\newcommand{\ket}[1]{\lvert #1\rangle}
\newcommand{\bra}[1]{\langle #1\rvert}
\newcommand{\braket}[2]{\langle #1\vert #2\rangle}
\newcommand{\proj}[1]{\lvert #1\rangle\langle #1\rvert}
\newcommand{\abs}[1]{\left\lvert #1\right\rvert}

\newcommand{\QFI}{\mathcal{H}}
\newcommand{\CFI}{\mathcal{F}}
\newcommand{\LQU}{\mathcal{U}}
\newcommand{\EOF}{\mathcal{E}}
\newcommand{\QD}{\mathcal{D}}
\newcommand{\MI}{\mathcal{I}}
\newcommand{\CC}{\mathcal{J}}
\newcommand{\hbin}{h_2}
\newcommand{\dmsq}{\Delta m^2}
\newcommand{\dmij}[2]{\Delta m^2_{#1#2}}
\newcommand{\etap}{\eta_{\mathrm{p}}}
\newcommand{\etam}{\eta_{\mathrm{m}}}

\newcommand{\xim}{\xi_{\mathrm{m}}}
\renewcommand{\Re}{\mathrm{Re}}
\renewcommand{\Im}{\mathrm{Im}}

\theoremstyle{plain}

\theoremstyle{definition}

\usepackage{tikz}
\usetikzlibrary{positioning,arrows.meta,calc,fit,backgrounds}
\usepackage{xcolor}

\definecolor{NJPBlue}{HTML}{1F4E79}
\definecolor{NJPBlueLight}{HTML}{EAF2F8}
\definecolor{NJPGray}{HTML}{F5F5F5}
\definecolor{NJPText}{HTML}{1F1F1F}

\begin{document}
\title{Quantum Estimation under Decoherence in Neutrino Oscillations: Quantum Resources, Flavor Accessibility, and Multiparameter Incompatibility}

\author{Jilali Loulijat\orcidlink{0009-0005-2517-7878}}
\affiliation{LPMS, Faculty of Sciences, Ibn Tofail University, Kenitra, Morocco.}\author{Abdallah Slaoui \orcidlink{0000-0002-5284-3240}}
\email{abdallah.slaoui@um5s.net.ma}
\affiliation{LPHE--Modeling and Simulation, Faculty of Sciences, Mohammed V University in Rabat, Rabat, Morocco.}\affiliation{Centre of Physics and Mathematics, CPM, Faculty of Sciences, Mohammed V University in Rabat, Rabat, Morocco.}\author{Mohamed Gouighri \orcidlink{0000-0002-9551-0251}}
\affiliation{LPMS, Faculty of Sciences, Ibn Tofail University, Kenitra, Morocco.}\author{Berihu Teklu \orcidlink{0000-0001-9280-533X}}
\email{berihu.gebrehiwot@ku.ac.ae}
\affiliation{College of Computing and Mathematical Sciences, Department of Applied Mathematics and Sciences and Center for Cyber-Physical Systems (C2PS), Khalifa University of Science and Technology, 127788, Abu Dhabi, United Arab Emirates.}


\begin{abstract}
Neutrino oscillations provide a natural interferometric setting in which open-system dynamics, quantum resources, and parameter-estimation limits can be studied within one framework. We distinguish two statistical families that are often conflated. In propagation-basis dephasing, a survival factor $\etap$ damps interference before flavor projection and therefore changes measurable flavor probabilities. In effective flavor-mode dephasing, a factor $\etam$ suppresses only the off-diagonal coherence of a single-particle, two-mode state while leaving its mode populations fixed. For the two-flavor mode state, we derive exact expressions for concurrence, entanglement of formation, and local quantum uncertainty and reduce one-sided projective quantum discord to a one-parameter optimization. In the regular interior $0<P<1$ and $0<\etam<1$, the QFIM in $(P,\chi,\etam)$ is diagonal and obeys $\QFI^{\rm mode}_{\chi\chi}=C_{\etam}^{2}$ and $\QFI^{\rm mode}_{\xim\xim}=C_{\etam}^{2}/(1-\etam^{2})$, where $\xim=-\ln\etam$. For propagation-basis dephasing, basis-matched QFI and flavor FI separate dynamical information loss from measurement inaccessibility for the mixing angle, mass-squared splitting, and dephasing rate. A monochromatic three-flavor benchmark then evaluates the mixed-state QFIM, flavor FIM, and SLD incompatibility for $(\theta_{23},\delta_{\rm CP},\xi_{31},g)$ at DUNE-like, T2HK-like, and ESSnuSB-inspired points. At $\Gamma_0=10^{-23}\,\mathrm{GeV}$, the fixed-coordinate ratio $\CFI^{\rm flav}_{\delta\delta}/\QFI_{\delta\delta}$ is approximately $0.010$, $0.013$, and $0.105$, respectively, while all three points exhibit strong $\theta_{23}$--$\xi_{31}$ incompatibility. Because one single-energy flavor measurement supplies only two independent probabilities, its four-parameter FIM is rank deficient; the quoted diagonal ratios are therefore conditional diagnostics rather than a joint four-parameter sensitivity. These results are state-level information-geometric benchmarks, not event-level sensitivity forecasts.
\end{abstract}

\maketitle

\section{Introduction}
Neutrino oscillations arise because weak-interaction flavor states are coherent superpositions of nondegenerate mass eigenstates~\cite{Pontecorvo1957,Pontecorvo1958,Maki1962,Bilenky2010,GiuntiKim}. Their relative propagation phases encode the mixing angles, mass-squared splittings, mass ordering, and the Dirac phase $\delta_{\rm CP}$. Reactor, atmospheric, and long-baseline experiments therefore act as interferometers whose precision depends both on the strength of the parameter encoding and on the survival of quantum coherence.

Open-system descriptions provide a controlled phenomenological language for departures from unitary propagation. Lindblad-type damping can represent stochastic matter fluctuations, environmental scattering, or effective quantum-spacetime noise, and it can both imitate nonstandard dynamics and degrade the extraction of standard oscillation parameters~\cite{Gorini1976,Lindblad1976,BenattiFloreanini2005,DeGouvea2020,DeGouvea2021,DeRomeri2023,Barenboim2024,Nandi2026,IceCube2024,Stankevich2026}. The operational effect depends on where the channel acts: propagation-basis dephasing changes flavor probabilities, whereas dephasing applied only to an effective flavor-mode coherence can leave the diagonal populations unchanged.

Quantum estimation theory provides the natural framework for separating these effects. The QFI quantifies the local distinguishability intrinsic to a state family, while the FI quantifies the information extracted by a specified measurement~\cite{Cramer,Rao,Helstrom,Holevo,BraunsteinCaves,ParisReview}. In multiparameter problems, a large QFIM is not by itself sufficient because the measurements that optimize different coordinates may be incompatible~\cite{Ragy2016,Candeloro,Bakmou2019,Asjad,BelliardoGiovannetti2021,Albarelli2019,Xia2023}. Recent neutrino studies have emphasized flavor-readout optimality, parameter-dependent basis choices, probability degeneracies, and multiparameter state geometry~\cite{FrugiueleParis2026,Yadav2026,HuangOhlssonVihonenZhou2026,YadavSubbaShiQFIMDegeneracy2026,ChundawatLi2026}.

The remaining gap is that quantum-resource analyses and neutrino-estimation analyses have largely been developed as separate narratives. A basis-explicit framework is still needed to distinguish four mechanisms that can produce a loss of precision: reduced distinguishability of the propagated state, information retained in the state but missed by flavor projection, decay of off-diagonal quantum resources, and incompatibility of the measurements that optimize different parameters. This distinction is especially important when the basis transformation or the noise basis itself depends on the parameter being estimated.

The present work extends earlier two-flavor resource analyses, including our own study~\cite{ElBouzaidi2025,LoulijatSciRep2026}. Its intended novelty is not another comparison of correlation measures alone, but their integration with exact resource--QFI identities, basis-matched QFI--FI comparisons, and a three-flavor mixed-state multiparameter benchmark. The central question is therefore operational: when decoherence reduces precision, can the loss be assigned to the state, to the available flavor measurement, to the disappearance of mode coherence, or to joint-estimation incompatibility?

The contributions are fourfold. First, propagation-basis and flavor-mode dephasing are kept as distinct statistical models, with separate survival factors and explicit statements of what population measurements can estimate. Second, the complete mode-state QFIM in $(P,\chi,\etam)$ is derived and related exactly to concurrence, entanglement of formation, and local quantum uncertainty. Third, basis-matched QFI and flavor FI are obtained for the effective mixing angle, mass-squared splitting, and dephasing strength. Fourth, a three-flavor benchmark compares state information, flavor accessibility, and SLD incompatibility for DUNE-like and T2HK-like constant-density configurations and an ESSnuSB-inspired second-maximum vacuum point.

The numerical analysis is deliberately restricted to monochromatic, fixed-coordinate, state-level diagnostics. It is not a global fit and does not include fluxes, cross sections, energy migration, detector response, backgrounds, priors, combined neutrino and antineutrino data, or nuisance-parameter marginalization. Moreover, a single flavor-probability vector has only two independent components and cannot jointly identify the four coordinates used in the benchmark. Within this restricted scope, the calculation exposes a pronounced measurement-accessibility gap for $\delta_{\rm CP}$ and a strong $\theta_{23}$--$\xi_{31}$ incompatibility, motivating an energy-binned analysis based on an attainable multiparameter bound.

The paper is organized as follows. Section~\ref{sec:model} introduces the oscillation model and the effective mode mapping. Section~\ref{sec:decoherence} defines the two dephasing families. Section~\ref{sec:resources} derives the quantum-resource diagnostics. Section~\ref{sec:qet} develops the QFI and FI results. Section~\ref{sec:results} presents the numerical protocol and benchmarks. Section~\ref{sec:discussion} discusses interpretation, limitations, and required extensions. Detailed derivations are collected in the appendices.
\FloatBarrier
\section{Oscillation model and effective two-qubit mapping}
\label{sec:model}

\subsection{Three-flavor vacuum evolution}

Let $\ket{\nu_\alpha}$, $\alpha=e,\mu,\tau$, denote the flavor basis and $\ket{\nu_i}$, $i=1,2,3$, the mass basis.  We use
\begin{equation}
    \ket{\nu_\alpha}=\sum_{i=1}^3 U_{\alpha i}^*\ket{\nu_i},
    \qquad
    \ket{\nu_i}=\sum_{\alpha=e,\mu,\tau} U_{\alpha i}\ket{\nu_\alpha},
    \label{eq:flavor_mass}
\end{equation}
where $U$ is the PMNS matrix~\cite{Bilenky2010}.  For Dirac neutrinos, in the standard convention,
\begin{widetext}
\begin{equation}
U=\begin{pmatrix}
 c_{12}c_{13} & s_{12}c_{13} & s_{13}\ee^{-\ii\delta_{\rm CP}} \\
 -s_{12}c_{23}-c_{12}s_{23}s_{13}\ee^{\ii\delta_{\rm CP}} &
  c_{12}c_{23}-s_{12}s_{23}s_{13}\ee^{\ii\delta_{\rm CP}} &
  s_{23}c_{13} \\
 s_{12}s_{23}-c_{12}c_{23}s_{13}\ee^{\ii\delta_{\rm CP}} &
 -c_{12}s_{23}-s_{12}c_{23}s_{13}\ee^{\ii\delta_{\rm CP}} &
 c_{23}c_{13}
\end{pmatrix},
\label{eq:pmns}
\end{equation}
\end{widetext}
with $c_{ij}=\cos\theta_{ij}$ and $s_{ij}=\sin\theta_{ij}$.  In the ultra-relativistic limit, after dropping a common phase, a neutrino produced as flavor $\alpha$ evolves as
\begin{equation}
    \ket{\nu_\alpha(L)}=\sum_{i=1}^3 U_{\alpha i}^*\exp\left[-\ii\frac{m_i^2L}{2E}\right]\ket{\nu_i}.
    \label{eq:three_state}
\end{equation}
The transition amplitude and probability are
\begin{align}
    \mathcal{A}_{\alpha\to\beta}(L,E)&=\sum_{i=1}^3 U_{\beta i}U_{\alpha i}^*\exp\left[-\ii\frac{m_i^2L}{2E}\right],
    \\
    P_{\alpha\to\beta}(L,E)&=\abs{\mathcal{A}_{\alpha\to\beta}(L,E)}^2.
\end{align}
Equivalently,
\begin{equation}
\begin{aligned}
P_{\alpha\to\beta}={}&\delta_{\alpha\beta}\\
&-4\sum_{i>j}\Re\left(U_{\alpha i}^*U_{\beta i}U_{\alpha j}U_{\beta j}^*\right)\sin^2\Delta_{ij}\\
&+2\sum_{i>j}\Im\left(U_{\alpha i}^*U_{\beta i}U_{\alpha j}U_{\beta j}^*\right)\sin 2\Delta_{ij}.
\end{aligned}
\label{eq:prob_3flavor}
\end{equation}
where
\begin{equation}
    \Delta_{ij}=\frac{\dmij{i}{j}L}{4E},
    \qquad
    \dmij{i}{j}=m_i^2-m_j^2.
\end{equation}
The CP-odd term is controlled by the imaginary rephasing invariants and vanishes in two-flavor limits.  This is why the CP phase must be treated in a genuine three-flavor setting, whereas the analytic correlation measures below can be developed in two-flavor reductions.

\subsection{Controlled two-flavor reduction}

In many experimentally relevant windows, one oscillation frequency dominates and the dynamics can be represented by an effective two-flavor model.  Let $\theta$ be the relevant effective mixing angle and let $\dmsq$ be the associated mass-squared splitting.  The flavor and mass bases are related by
\begin{equation}
\begin{pmatrix}
\ket{\nu_\alpha}\\
\ket{\nu_\beta}
\end{pmatrix}
=
\begin{pmatrix}
\cos\theta & \sin\theta\\
-\sin\theta & \cos\theta
\end{pmatrix}
\begin{pmatrix}
\ket{\nu_i}\\
\ket{\nu_j}
\end{pmatrix}.
\label{eq:two_rotation}
\end{equation}
After removing a common phase, define
\begin{equation}
    \Phi=\frac{\dmsq L}{2E},
    \qquad
    \Delta=\frac{\Phi}{2}=\frac{\dmsq L}{4E}.
    \label{eq:oscillation_phase}
\end{equation}
The propagated state in the mass basis is
\begin{equation}
    \ket{\psi(L)}=\cos\theta\ket{\nu_i}+\sin\theta\ee^{-\ii\Phi}\ket{\nu_j}.
    \label{eq:two_mass_state}
\end{equation}
Projecting back to the flavor basis gives
\begin{equation}
    \ket{\psi(L)}=A_{\alpha\alpha}\ket{\nu_\alpha}+A_{\alpha\beta}\ket{\nu_\beta},
    \label{eq:two_flavor_state}
\end{equation}
with
\begin{align}
    A_{\alpha\alpha}&=\cos^2\theta+\sin^2\theta\ee^{-\ii\Phi},
    \\
    A_{\alpha\beta}&=\sin\theta\cos\theta\left(\ee^{-\ii\Phi}-1\right).
\end{align}
Thus
\begin{align}
    P_{\alpha\to\beta}&=\abs{A_{\alpha\beta}}^2=\sin^2(2\theta)\sin^2\Delta,
    \label{eq:two_prob_trans}
    \\
    P_{\alpha\to\alpha}&=1-P_{\alpha\to\beta}.
    \label{eq:two_prob_surv}
\end{align}
We write
\begin{equation}
    P\equiv P_{\alpha\to\beta},
    \qquad
    S\equiv P_{\alpha\to\alpha}=1-P.
\end{equation}

The two-flavor approximation is not a substitute for precision global fits.  In a three-flavor treatment, KamLAND-like reactor propagation receives a leading factor $c_{13}^4$ and a small $s_{13}^4$ correction, Daya Bay-like disappearance contains the effective splitting $\Delta m^2_{ee}$ plus a subleading solar contribution, and accelerator disappearance is affected by $s_{13}^2$, $\dmij{2}{1}/\dmij{3}{1}$, and matter corrections.  These effects shift phases and amplitudes but do not change the analytic role of coherence in the two-level subsystem.

\subsection{Mode-entangled representation}
The occupation-number mapping follows the standard single-particle
mode-entanglement construction used in quantum-information analyses of
neutrino oscillations \cite{Blasone2009,Blasone2013,Alok2016,AlokBlasone2025}. The two
qubits label occupation modes associated with the two effective flavor states;
they do not represent two independently addressable neutrino particles. EOF,
projective QD, and LQU are therefore used below as diagnostics of coherent mode
superposition and measurement-induced quantumness, not as claims of directly
extractable nonlocal entanglement between two particles.

In the effective-mode representation, the two-dimensional flavor state may be embedded in a two-qubit mode basis.  We identify
\begin{equation}
    \ket{\nu_\alpha}\equiv\ket{01},
    \qquad
    \ket{\nu_\beta}\equiv\ket{10},
\end{equation}
so that
\begin{equation}
    \ket{\psi(L)}=A_{\alpha\alpha}\ket{01}+A_{\alpha\beta}\ket{10}.
    \label{eq:mode_state}
\end{equation}
This is a single-particle mode-entangled state.  The density matrix in the ordered basis $\{\ket{00},\ket{01},\ket{10},\ket{11}\}$ is
\begin{equation}
\rho_0=
\begin{pmatrix}
0&0&0&0\\
0&S&z&0\\
0&z^*&P&0\\
0&0&0&0
\end{pmatrix},
\qquad
z=A_{\alpha\alpha}A_{\alpha\beta}^*.
\label{eq:rho_pure_block}
\end{equation}
For the pure state, $\abs{z}=\sqrt{SP}$.  The diagonal entries carry the flavor probabilities; the off-diagonal element $z$ carries the coherence between the two mode occupations.

\FloatBarrier
\section{Open-system description of decoherence}
\label{sec:decoherence}

\subsection{Lindblad evolution}

An isolated neutrino evolves according to $\dot{\rho}=-\ii[H,\rho]$.  A Markovian open quantum system is instead described by the Gorini-Kossakowski-Sudarshan-Lindblad equation
\begin{equation}
    \frac{\dd\rho}{\dd t}=-\ii[H,\rho]+
    \sum_k\left(L_k\rho L_k^\dagger-\frac{1}{2}\{L_k^\dagger L_k,\rho\}\right).
    \label{eq:lindblad}
\end{equation}
Complete positivity and trace preservation are built into this form.  If the Lindblad operators are Hermitian, the von Neumann entropy is nondecreasing.  In neutrino phenomenology, the dissipative term may represent stochastic matter profiles, fluctuating fields, radiative effects, or quantum-spacetime noise~\cite{BenattiFloreanini2005,DeGouvea2020,DeGouvea2021,DeRomeri2023,Nandi2026}.  The analysis of quantum decoherence effects at DUNE and T2HK emphasizes that such damping can both mimic and degrade the extraction of standard oscillation parameters, and that combined baselines help break degeneracies.

For three flavors it is common to expand $\rho$ in the Gell--Mann basis, $\rho=I/3+\frac{1}{2}\sum_{a=1}^8\rho_a\lambda_a$, so that $\dot{\bm\rho}=(M+M_D)\bm\rho$. As an example used in long-baseline phenomenology, one may write
\begin{equation}
    M_D=-\diag(\Gamma_1,\Gamma_2,\Gamma_1,\Gamma_1,\Gamma_2,\Gamma_1,\Gamma_2,\Gamma_1),
    \label{eq:gellmann_dissipator}
\end{equation}
which is a symmetry-reduced ansatz rather than a unique consequence of the Lindblad equation~\cite{Barenboim2024}. Because diagonal Gell--Mann components encode population differences, Eq.~\eqref{eq:gellmann_dissipator} is not automatically equivalent to pure dephasing that damps only off-diagonal propagation-basis density-matrix elements. Microscopic fluctuating-matter or scattering models can generate independent rates and more general dissipative coefficients~\cite{BenattiFloreanini2005,Stankevich2026}. In this manuscript, Eq.~\eqref{eq:gellmann_dissipator} should be retained only as literature context; the actual three-flavor numerical channel is defined operationally by Eq.~\eqref{eq:threeflavor_dephased_state} below.
\subsection{Propagation-basis dephasing and damped flavor probabilities}

The minimal two-flavor propagation-basis model is
\begin{equation}
    \rho_m(L)=
    \begin{pmatrix}
    \cos^2\theta & \etap\sin\theta\cos\theta\ee^{\ii\Phi}\\
    \etap\sin\theta\cos\theta\ee^{-\ii\Phi} & \sin^2\theta
    \end{pmatrix},
    \label{eq:mass_dephasing}
\end{equation}
where
\begin{equation}
    \etap(L,E)=\exp[-\Gamma_{\mathrm p}(E)L].
    \label{eq:damping_factor}
\end{equation}
Flavor projection gives
\begin{equation}
    P_{\alpha\to\beta}^{\rm prop}(L,E;\Gamma_{\mathrm p})
    =\frac{1}{2}\sin^2(2\theta)
    \left[1-\etap(L,E)\cos\Phi\right].
    \label{eq:damped_probability}
\end{equation}
The unitary expression is recovered at $\etap=1$, while $\etap\to0$ gives the
incoherent average $\frac{1}{2}\sin^2(2\theta)$. Because $\etap$ acts before
flavor projection, the damping changes the measured populations and can in
principle be inferred from flavor counting, except at phases where the
probability derivative with respect to the rate vanishes.

\subsection{Flavor-mode dephasing of the effective two-qubit state}

The resource analysis uses a different statistical family. The mode
populations $S$ and $P$ of Eq.~\eqref{eq:rho_pure_block} are held fixed, while
the coherence between the two occupation modes is reduced according to
\begin{equation}
\rho_{\etam}=
\begin{pmatrix}
0&0&0&0\\
0&S&\etam z&0\\
0&\etam z^*&P&0\\
0&0&0&0
\end{pmatrix},
\qquad 0\leq \etam\leq 1.
\label{eq:rho_eta}
\end{equation}
Here $\etam$ is a mode-coherence survival factor; it is not identified with
$\etap$ unless an explicit model is supplied that maps the two channels onto
one another. A measurement of the diagonal mode populations cannot estimate
$\etam$, because the probability distribution $(S,P)$ is independent of this
parameter. The nonzero eigenvalues are
\begin{equation}
    \lambda_{\pm}=\frac{1\pm R}{2},
    \qquad
    R=\sqrt{(1-2P)^2+4\etam^2P(1-P)}.
    \label{eq:eigs_rho_eta}
\end{equation}
This rank-two form is the starting point for the exact resource expressions
and the mode-state QFIM derived below.

\begin{table*}
\caption{Operational distinction between the two dephasing models. Separate symbols are necessary because the channels act on different statistical families and need not represent the same microscopic process.}
\label{tab:dephasing_models}
\centering
\footnotesize
\begin{tabular}{@{}p{0.19\linewidth}p{0.25\linewidth}p{0.23\linewidth}p{0.25\linewidth}@{}}
\toprule
Model & Quantity damped & Effect on measured populations & Can population counting estimate the damping?\\
\midrule
Propagation-basis dephasing, Eq.~\eqref{eq:mass_dephasing}
& Interference between propagation eigenstates before flavor projection
& Flavor probabilities change through Eq.~\eqref{eq:damped_probability}
& Yes, except at points where the local probability derivative vanishes.\\
Flavor-mode dephasing, Eq.~\eqref{eq:rho_eta}
& Off-diagonal coherence of the effective single-particle, two-mode state
& Mode populations remain fixed at $(S,P)$
& No. A population-only measurement contains no information on $\etam$.\\
\bottomrule
\end{tabular}
\end{table*}

\section{Quantum-information resources}
\label{sec:resources}

\subsection{Entanglement of formation}

For a two-qubit state, Wootters' concurrence is~\cite{Wootters1998,Bennett1996}
\begin{equation}
    C(\rho)=\max\left\{0,\sqrt{\lambda_1}-\sqrt{\lambda_2}-\sqrt{\lambda_3}-\sqrt{\lambda_4}\right\},
\end{equation}
where $\lambda_i$ are the eigenvalues, in decreasing order, of $\rho(\sigma_y\otimes\sigma_y)\rho^*(\sigma_y\otimes\sigma_y)$.  For Eq.~\eqref{eq:rho_eta}, the X-state structure gives the exact result
\begin{equation}
    C_{\etam}=2\etam\sqrt{P(1-P)}.
    \label{eq:concurrence_eta}
\end{equation}
The entanglement of formation is
\begin{equation}
    \EOF(\rho_{\etam})=\hbin\left(\frac{1+\sqrt{1-C_{\etam}^2}}{2}\right),
    \label{eq:eof_eta}
\end{equation}
where
\begin{equation}
    \hbin(x)=-x\log_2x-(1-x)\log_2(1-x).
\end{equation}
Thus entanglement vanishes at $P=0,1$ and is maximal when $P=1/2$ and $\etam=1$.  A nonmaximal mixing angle restricts the accessible range of $P$ and therefore caps the maximum possible entanglement.

 The concurrence in Eq.~\eqref{eq:concurrence_eta} shows that entanglement is controlled by two independent ingredients: the population balance $P(1-P)$ and the coherence survival factor $\etam$. Even if the transition probability is sizable, entanglement can be strongly reduced when the dephasing channel suppresses the off-diagonal term. Therefore, EOF is a stricter witness of coherent oscillation quantumness than the transition probability itself. The maximum EOF is reached only when the effective flavor state is both balanced, $P=1/2$, and fully coherent, $\etam=1$.
\subsection{One-sided projective quantum discord}
Projective quantum discord quantifies nonclassical measurement disturbance that can survive in separable states~\cite{OllivierZurek2001,HendersonVedral2001,Modi2012,AliRauAlber2010,GirolamiAdesso2011}. The quantum mutual information is
\begin{equation}
    \MI(\rho)=S(\rho_A)+S(\rho_B)-S(\rho),
    \label{eq:mutual_info}
\end{equation}
where $S(\rho)=-\Tr(\rho\log_2\rho)$.  For Eq.~\eqref{eq:rho_eta}, the reduced states have eigenvalues $\{P,1-P\}$ and therefore
\begin{equation}
    \MI(\rho_{\etam})=2\hbin(P)-\hbin\left(\frac{1+R}{2}\right).
    \label{eq:mutual_eta}
\end{equation}
The one-sided classical correlation for measurements on subsystem $B$ is
\begin{equation}
    \CC_B(\rho)=S(\rho_A)-\min_{\{\Pi_k^B\}}\sum_k p_kS(\rho_{A|k}),
\end{equation}
with $p_k=\Tr[(I\otimes\Pi_k^B)\rho]$ and $\rho_{A|k}=\Tr_B[(I\otimes\Pi_k^B)\rho(I\otimes\Pi_k^B)]/p_k$.  The one-sided projective discord is
\begin{equation}
    \QD_B(\rho)=\MI(\rho)-\CC_B(\rho).
    \label{eq:discord_def}
\end{equation}
For the present rank-two X state, the minimization can be written as a one-parameter projective search.  Let
\begin{equation}
\begin{aligned}
\ket{0_\vartheta}&=\cos\frac{\vartheta}{2}\ket{0}
+\ee^{\ii\varphi}\sin\frac{\vartheta}{2}\ket{1},\\
\ket{1_\vartheta}&=\sin\frac{\vartheta}{2}\ket{0}
-\ee^{\ii\varphi}\cos\frac{\vartheta}{2}\ket{1}.
\end{aligned}
\end{equation}
The minimizing phase is aligned with $\arg z$, so only $\vartheta\in[0,\pi]$ must be scanned. Throughout this work, QD refers to this optimization over rank-one projective measurements on subsystem $B$, not to an unrestricted POVM optimization.  Defining $q_0$ and $q_1$ as the two outcome probabilities and $\bm r_0$, $\bm r_1$ as the corresponding conditional Bloch vectors of subsystem $A$, the conditional entropy is
\begin{equation}
    S_{\rm cond}(\vartheta)=\sum_{k=0}^1 q_k\hbin\left(\frac{1+\abs{\bm r_k}}{2}\right),
    \label{eq:discord_cond}
\end{equation}
so that
\begin{equation}
    \QD_B(\rho_{\etam})=\hbin(P)-\hbin\left(\frac{1+R}{2}\right)+\min_\vartheta S_{\rm cond}(\vartheta).
    \label{eq:discord_eta}
\end{equation}
For $\etam=1$, the total state is pure and this reduces to
\begin{equation}
    \QD_B(\rho_0)=\hbin(P),
\end{equation}
which equals the entanglement entropy.  For $0<\etam<1$, discord can persist in regions where concurrence and EOF are strongly suppressed, because discord captures nonclassical measurement disturbance beyond distillable entanglement.

\subsection{Local quantum uncertainty}

The local quantum uncertainty with respect to subsystem $A$ is the minimum Wigner-Yanase skew information~\cite{WignerYanase1963} over local observables with nondegenerate spectrum~\cite{Girolami2013}.  For two-qubit states it has the closed form
\begin{equation}
\begin{aligned}
\LQU_A(\rho)&=1-\lambda_{\max}(W),\\
W_{ij}&=\Tr\left[\sqrt{\rho}(\sigma_i\otimes I)\sqrt{\rho}(\sigma_j\otimes I)\right].
\end{aligned}
\label{eq:lqu_general}
\end{equation}
where $i,j=x,y,z$.  For Eq.~\eqref{eq:rho_eta}, one obtains
\begin{equation}
    \LQU_A(\rho_{\etam})=
    \frac{4\etam^2P(1-P)}{1+2\sqrt{P(1-P)(1-\etam^2)}}.
    \label{eq:lqu_eta}
\end{equation}
In the unitary limit,
\begin{equation}
    \LQU_A(\rho_0)=4P(1-P)=C_0^2.
    \label{eq:lqu_pure}
\end{equation}
Equation~\eqref{eq:lqu_eta} is one of the most useful compact results of the two-qubit analysis.  It shows that LQU is controlled by both the transition probability and the surviving coherence.  Since LQU is related to interferometric power and bounds a local quantum Fisher information, it provides a bridge between correlation diagnostics and quantum metrology.
\begin{figure*}
		{{\begin{minipage}[t]{.33\linewidth}
					\centering
					(a)\par\vspace{1mm}
\includegraphics[width=\linewidth]{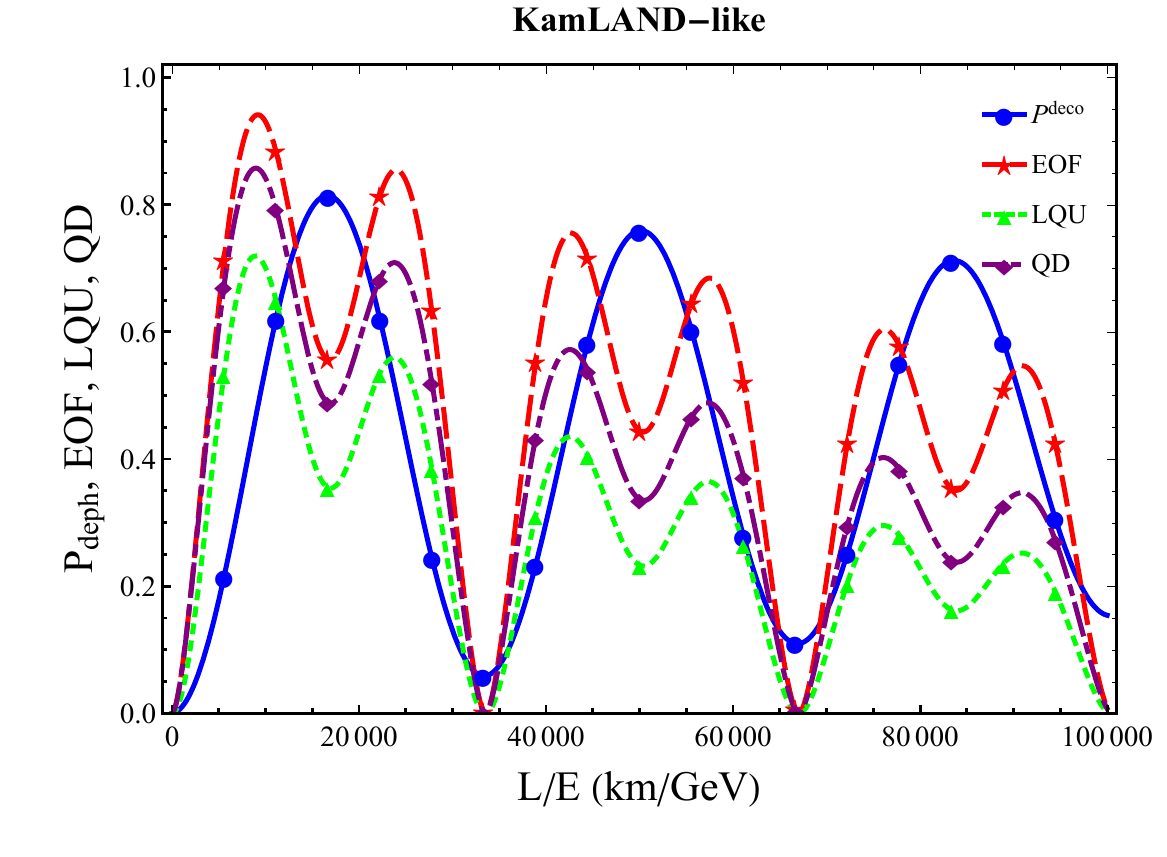}
				\end{minipage}\hfill
				\begin{minipage}[t]{.33\linewidth}
					\centering
					(b)\par\vspace{1mm}
\includegraphics[width=\linewidth]{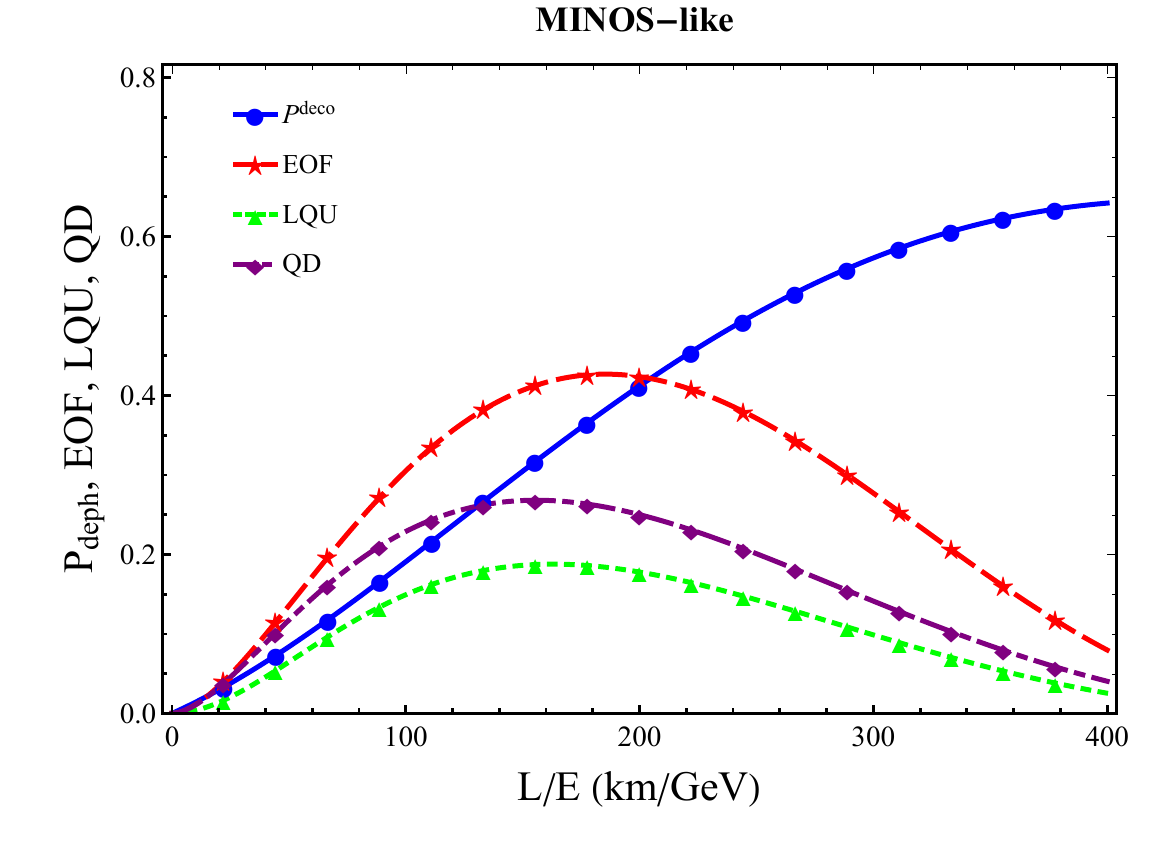}
		\end{minipage}}
	\begin{minipage}[t]{.33\linewidth}
		\centering
		(c)\par\vspace{1mm}
\includegraphics[width=\linewidth]{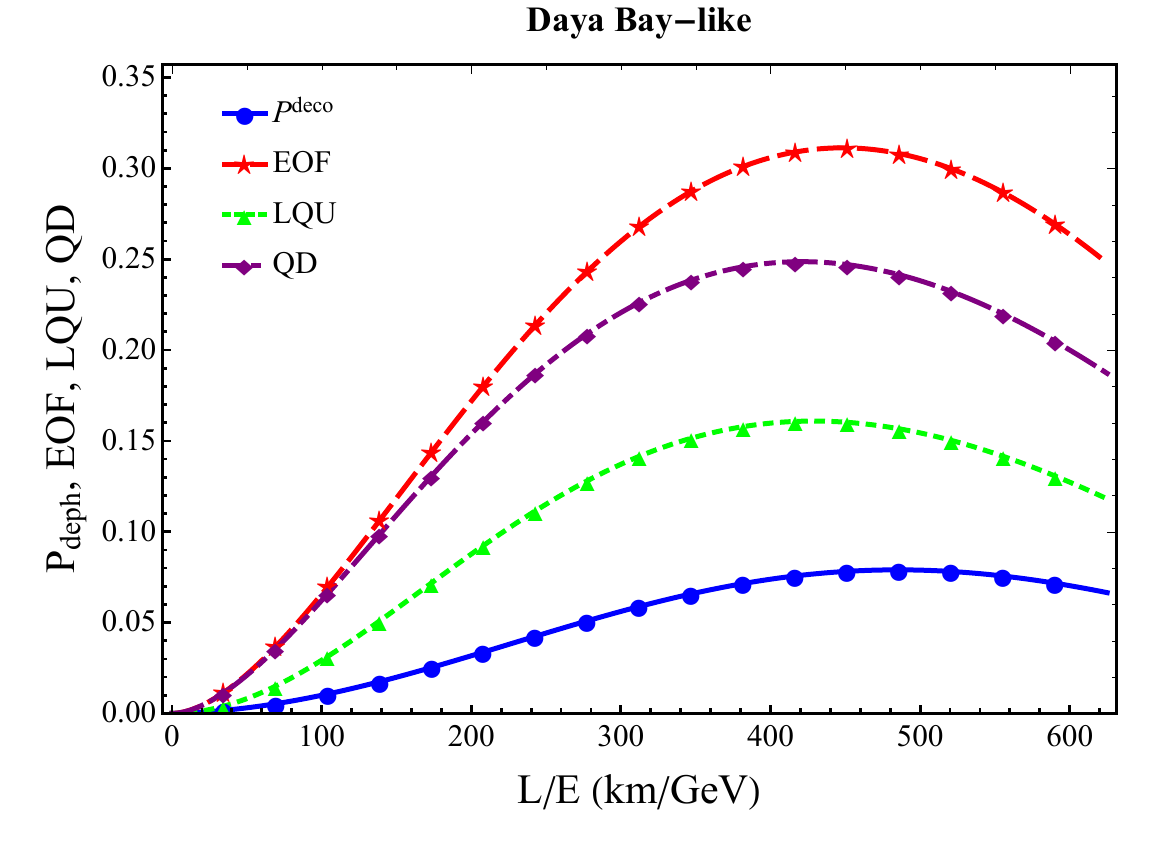}
\end{minipage}}
		\caption{Illustrative comparison of two operationally distinct responses in the KamLAND-, MINOS-, and Daya Bay-like regimes. The probability curve is computed from propagation-basis dephasing, Eq.~\eqref{eq:damped_probability}, whereas EOF, projective QD, and LQU are evaluated for the flavor-mode state in Eq.~\eqref{eq:rho_eta}. The same numerical damping profile is used only to compare decay scales; the curves do not imply that $\etap$ and $\etam$ define the same physical channel. Panels (a)--(c) are effective two-flavor diagnostics and do not contain the intrinsically three-flavor CP phase.}
        \label{fig:resources_vs_eta}
	\end{figure*}
As shown in Fig.~\ref{fig:resources_vs_eta}, the entanglement of formation, discord, and local quantum uncertainty track one another closely for the MINOS-like (near-maximal mixing) regime, whereas the Daya Bay-like regime never approaches the maximal-resource point because the accessible transition probability $P$ is bounded by the small value of $\theta_{13}$.

\section{Quantum estimation theory for neutrino oscillations}
\label{sec:qet}

\subsection{Classical and quantum Cram\'er--Rao bounds}

Let $x$ denote the outcome of a measurement whose conditional probability distribution is $p(x|\lambda)$.  For $M$ independent trials, the variance of any unbiased estimator $\tilde\lambda$ satisfies
\begin{equation}
    \Var(\tilde\lambda)\geq \frac{1}{M\CFI(\lambda)},
    \qquad
    \CFI(\lambda)=\sum_x\frac{[\partial_\lambda p(x|\lambda)]^2}{p(x|\lambda)}.
    \label{eq:classical_crb}
\end{equation}
In quantum mechanics, $p(x|\lambda)=\Tr[\rho_\lambda\Pi_x]$ for a POVM $\{\Pi_x\}$.  Optimizing over all allowed measurements gives the quantum Fisher information $\QFI(\lambda)$ and the quantum Cram'er--Rao bound
\begin{equation}
    \Var(\tilde\lambda)\geq \frac{1}{M\CFI(\lambda)}\geq \frac{1}{M\QFI(\lambda)}.
    \label{eq:qcrb}
\end{equation}
The QFI is defined by the symmetric logarithmic derivative $L_\lambda$,
\begin{equation}
    \partial_\lambda\rho_\lambda=\frac{1}{2}\left(\rho_\lambda L_\lambda+L_\lambda\rho_\lambda\right),
    \qquad
    \QFI(\lambda)=\Tr(\rho_\lambda L_\lambda^2).
\end{equation}
If $\rho_\lambda=\ket{\psi_\lambda}\bra{\psi_\lambda}$ is pure,
\begin{equation}
    \QFI(\lambda)=4\left(\braket{\partial_\lambda\psi_\lambda}{\partial_\lambda\psi_\lambda}
    -\abs{\braket{\psi_\lambda}{\partial_\lambda\psi_\lambda}}^2\right).
    \label{eq:pure_qfi}
\end{equation}
For a spectral decomposition $\rho_\lambda=\sum_n p_n\ket{n}\bra{n}$,
\begin{equation}
    \QFI(\lambda)=\sum_{p_n>0}\frac{(\partial_\lambda p_n)^2}{p_n}
    +2\sum_{m,n}\frac{(p_m-p_n)^2}{p_m+p_n}\abs{\braket{m}{\partial_\lambda n}}^2,
    \label{eq:spectral_qfi}
\end{equation}
where terms with $p_m+p_n=0$ are omitted.  The first term is population information and the second term is basis-rotation information.

\subsection{Physical basis and parameter-dependent encodings}
\label{subsec:basis_dependence}
The inequality $\CFI\leq\QFI$ compares a fixed statistical model $\rho_\lambda$ with a parameter-independent POVM. The QFI is invariant under a parameter-independent unitary change of representation, $\rho_\lambda\mapsto V\rho_\lambda V^\dagger$. If the unitary depends on the unknown parameter,
\begin{equation}
\rho_\lambda\mapsto V_\lambda\rho_\lambda V_\lambda^\dagger,
\end{equation}
then $\partial_\lambda V_\lambda$ contributes to the derivative and the transformed family generally describes a different encoding, not merely the same family in new coordinates. Likewise, if the POVM itself depends on $\lambda$, its derivative contributes to the outcome distribution and the standard Braunstein--Caves comparison must be reformulated. These distinctions are central in multiparameter metrology~\cite{Ragy2016,Candeloro,Asjad,BelliardoGiovannetti2021} and are especially relevant for neutrinos because the PMNS matrix contains the mixing angles and $\delta_{\rm CP}$.

Throughout the basis-matched comparisons below, the laboratory convention is that flavor outcomes are fixed parameter-independent detector labels. All dependence of PMNS-parameters is assigned to the preparation-and-propagation map that produces the state expressed in that fixed flavor basis. A mass basis family differentiated while holding the mass basis fixed is therefore a different statistical encoding from the corresponding family obtained after the parameter-dependent flavor rotation. Consequently, every reported QFI must be read together with its preparation map, Hilbert-space basis, noise basis, and accessible measurement.

\subsection{Pure two-flavor QFI}

For the mass-basis of two-flavors in Eq.~\eqref{eq:two_mass_state}, the QFI with respect to the mixing angle is
\begin{equation}
    \QFI(\theta)=4.
    \label{eq:qfi_theta_two}
\end{equation}
The QFI with respect to the relative phase $\Phi$ is
\begin{equation}
    \QFI(\Phi)=\sin^2(2\theta).
    \label{eq:qfi_phi_two}
\end{equation}
Therefore, by the chain rule,
\begin{align}
    \QFI(\dmsq)&=\sin^2(2\theta)\left(\frac{L}{2E}\right)^2,
    \label{eq:qfi_dmsq_two}\\
    \QFI(L)&=\sin^2(2\theta)\left(\frac{\dmsq}{2E}\right)^2,
    \\
    \QFI(E)&=\sin^2(2\theta)\left(\frac{\dmsq L}{2E^2}\right)^2.
\end{align}
These expressions are state-level bounds.  They do not yet specify whether flavor measurements attain them.
\subsection{Flavor-measurement FI and saturation}
For two outcomes $\{\alpha,\beta\}$, the FI for a parameter $\lambda$ is
\begin{equation}
    \CFI_{\rm flav}(\lambda)=\frac{[\partial_\lambda P(\lambda)]^2}{P(\lambda)[1-P(\lambda)]}.
    \label{eq:fi_two_outcome}
\end{equation}
For phase estimation using $P=A\sin^2(\Phi/2)$ with $A=\sin^2(2\theta)$,
\begin{equation}
    \CFI_{\rm flav}(\Phi)=
    \frac{A^2\sin^2\Phi}{4A\sin^2(\Phi/2)\left[1-A\sin^2(\Phi/2)\right]}.
    \label{eq:fi_phase_flav}
\end{equation}
The efficiency ratio is
\begin{equation}
    \mathcal{R}_\Phi=\frac{\CFI_{\rm flav}(\Phi)}{\QFI(\Phi)}
    =\frac{\cos^2(\Phi/2)}{1-A\sin^2(\Phi/2)}.
    \label{eq:ratio_phase}
\end{equation}
For maximal mixing, $A=1$, flavor measurement saturates the QFI for all nonsingular phases.  For nonmaximal mixing, saturation occurs near $\Phi=0$ modulo $2\pi$ while efficiency can be poor near oscillation maxima.  This simple result illustrates why optimality depends on both the parameter and the location in $L/E$.

For mixing-angle estimation,
\begin{equation}
    \partial_\theta P=2\sin(4\theta)\sin^2\frac{\Phi}{2},
\end{equation}
so that
\begin{equation}
    \CFI_{\rm flav}(\theta)=
    \frac{4\sin^2(4\theta)\sin^4(\Phi/2)}{A\sin^2(\Phi/2)[1-A\sin^2(\Phi/2)]}.
    \label{eq:fi_theta_flav}
\end{equation}
At the first oscillation maximum, $\Phi=\pi$, this becomes
\begin{equation}
    \CFI_{\rm flav}(\theta)=16,
\end{equation}
whenever $\theta$ is not exactly maximal. This value exceeds the mass-basis-fixed QFI of Eq.~\eqref{eq:qfi_theta_two}, $\QFI^{(m)}(\theta)=4$, which only appears to violate the quantum Cram\'er-Rao bound $\CFI\le\QFI$: the two objects are evaluated for different encodings of the same physical parameter, exactly the subtlety flagged in Sec.~\ref{subsec:basis_dependence}. The QFI that is directly comparable to a flavor measurement is the one obtained after the $\theta$-dependent rotation to the fixed flavor basis is taken into account, $\QFI^{(f)}(\theta)$, derived in Sec.~\ref{subsec:qfi-fi-mass-dephasing} [Eq.~\eqref{eq:qfi-theta-dephasing}], which gives $\QFI^{(f)}(\theta)=16$ at $\etap=1$, $\Phi=\pi$. With this basis-matched comparison, flavor measurement saturates the two-flavor QFI for the mixing angle at the first maximum, $\CFI_{\rm flav}(\theta)=\QFI^{(f)}(\theta)=16$. This analytic observation is the two-flavor counterpart of the broader result emphasized by Paris and collaborators for several PMNS mixing angles.

\begin{figure}[!htbp]
\centering
\includegraphics[width=0.8\linewidth]{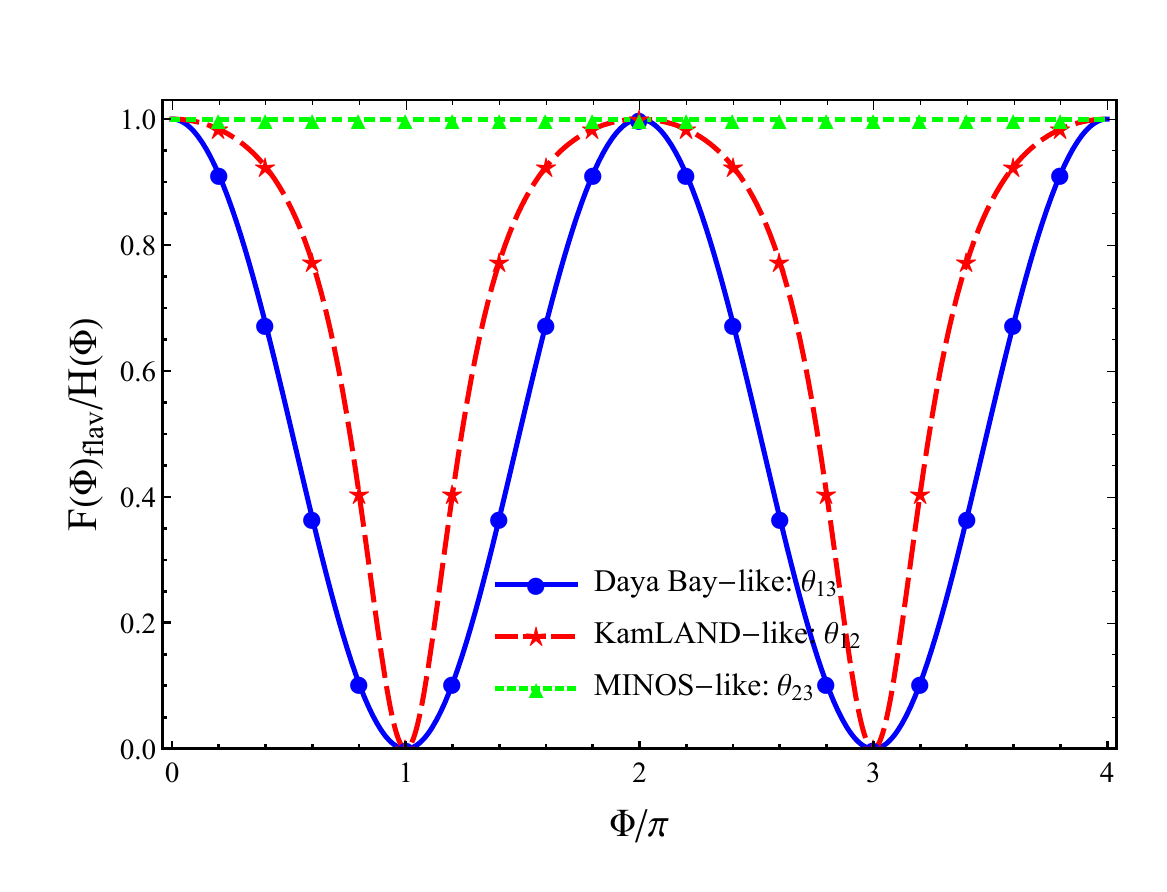}
\caption{Flavor-measurement efficiency $\CFI_{\rm flav}(\Phi)/\QFI(\Phi)$ for two-flavor phase estimation.  The efficiency depends strongly on the effective mixing amplitude $A=\sin^2(2\theta)$ and on the oscillation phase.  The plot illustrates why a flavor projection can be optimal in some regimes and inefficient in others.}
\label{fig:fisher_efficiency}
\end{figure}

\subsection{Three-flavor QFI for PMNS parameters}

For a three-flavor accelerator beam produced as $\nu_\mu$, the state is
\begin{equation}
    \ket{\psi_\mu(L;\bm\lambda)}=\sum_{i=1}^3 U_{\mu i}^*(\bm\lambda)\ee^{-\ii m_i^2L/(2E)}\ket{\nu_i},
\end{equation}
where $\bm\lambda=(\theta_{12},\theta_{13},\theta_{23},\delta_{\rm CP},\dmij{2}{1},\dmij{3}{1})$.For any pair of parameters $\lambda_a$ and $\lambda_b$, the pure-state
quantum Fisher information matrix is
\begin{equation}
    \QFI_{ab}=4\Re\left[\braket{\partial_a\psi}{\partial_b\psi}
    -\braket{\partial_a\psi}{\psi}\braket{\psi}{\partial_b\psi}\right]
    \label{eq:qfi_matrix_pure}
\end{equation}
defines the multiparameter QFI matrix.  The single-parameter QFI is the diagonal element if all other parameters are fixed.  Flavor detection gives
\begin{equation}
    \CFI_{ab}^{\rm flav}=\sum_{\beta=e,\mu,\tau}\frac{\partial_a P_{\mu\to\beta}\,\partial_b P_{\mu\to\beta}}{P_{\mu\to\beta}}.
    \label{eq:cfi_matrix_flavor}
\end{equation}
These expressions form the computational core of the Paris analysis~\cite{FrugiueleParis2026} and are also the starting point for the recent flavor-basis multiparameter QFIM studies~\cite{HuangOhlssonVihonenZhou2026,YadavSubbaShiQFIMDegeneracy2026}.  The essential conclusions can be stated operationally.  Mixing angles are amplitude-like parameters and can be optimally accessed by flavor projection near appropriate oscillation maxima.  The CP phase is a phase-like parameter in a genuinely three-flavor interference term; flavor measurements extract only part of the available information, especially near the first maximum, while the second maximum enhances relative CP sensitivity.
\subsection{How the present QFIM use differs from degeneracy lifting}
The multiparameter QFIM in Eq.~\eqref{eq:qfi_matrix_pure} can be used in two operationally distinct ways.  In degeneracy studies, as in Ref.~\cite{YadavSubbaShiQFIMDegeneracy2026}, two parameter sets are chosen so that the measured appearance probability is the same, and one asks whether the fidelity or the element-wise QFIM difference $\Delta F_{ij}$ reveals that the underlying quantum states are nevertheless distinct.  In the present work, the purpose is different.  We ask how decoherence changes the quantum resources that make such distinguishability possible in the first place, and how much of the remaining state information can be accessed by the flavor measurement actually available in oscillation experiments.  Thus the QFIM is used here as a quantum-sensing benchmark for resource degradation, not as a standalone degeneracy-resolution diagnostic.

This distinction is important for interpreting claims of novelty.  Ref.~\cite{YadavSubbaShiQFIMDegeneracy2026} establishes that a probability degeneracy in $(\theta_{23},\delta_{\rm CP})$ can be lifted by the geometry of the three-flavor state.  Our extension instead shows that decoherence can reduce precisely the off-diagonal coherence and nonclassical correlations that feed the state geometry.  In this sense, decoherence is both a possible new-physics signal and a metrological nuisance: it can be estimated as an open-system parameter, but it can also compress the QFI landscape and weaken the resource-based distinction between otherwise nearby oscillation scenarios.
\subsection{Exact mode-state QFIM under dephasing}
\label{subsec:mode_qfim}
Any rank-two density matrix can be written as
\begin{equation}
    \rho_{\boldsymbol\lambda}=\frac{1}{2}\left(I+\bm r_{\boldsymbol\lambda}\cdot\bm\sigma\right),
    \label{eq:bloch_representation}
\end{equation}
where, for $|\bm r|<1$,
\begin{equation}
    \QFI_{ab}=\partial_a\bm r\cdot\partial_b\bm r+
    \frac{(\bm r\cdot\partial_a\bm r)(\bm r\cdot\partial_b\bm r)}{1-|\bm r|^2}.
    \label{eq:qubit_qfi}
\end{equation}
For Eq.~\eqref{eq:rho_eta}, write $z=\sqrt{P(1-P)}e^{\ii\chi}$. The Bloch vector is
\begin{equation}
\bm r=\begin{pmatrix}
2\etam\sqrt{P(1-P)}\cos\chi\\
2\etam\sqrt{P(1-P)}\sin\chi\\
1-2P
\end{pmatrix}^{\!T}.
\label{eq:bloch_eta}
\end{equation}
Direct substitution into Eq.~\eqref{eq:qubit_qfi} gives the exact diagonal QFIM
\begin{equation}
\begin{aligned}
&\QFI^{\rm mode}(P,\chi,\etam)\\
&\quad=\diag\!\left[
\frac{1}{P(1-P)},\;
4\etam^2P(1-P),\;
\frac{4P(1-P)}{1-\etam^2}
\right].
\end{aligned}
\label{eq:mode_qfim}
\end{equation}
Equation~\eqref{eq:mode_qfim} applies in the regular interior of the statistical model, $0<P<1$ and $0<\etam<1$. At $P=0,1$ or $\etam=1$, the rank or support changes and the differentiability and constant-rank assumptions behind the local SLD Cram\'er--Rao analysis fail. The limiting QFI may then be discontinuous and should not be interpreted as an unconstrained experimental precision~\cite{Safranek2017}.

All cross terms vanish. The first entry is population information, the second
is coherent-phase information, and the third is information on the dephasing
amplitude. Since the concurrence is
$C_{\etam}=2\etam\sqrt{P(1-P)}$, one obtains
\begin{equation}
\QFI^{\rm mode}_{\chi\chi}=C_{\etam}^2,
\qquad
\LQU_A=\frac{\QFI^{\rm mode}_{\chi\chi}}
{1+2\sqrt{P(1-P)(1-\etam^2)}}.
\label{eq:resource_qfi_identities}
\end{equation}
Thus EOF is a monotonic function of $\sqrt{\QFI^{\rm mode}_{\chi\chi}}$, whereas
LQU is a mixedness-dependent fraction of the phase QFI.

For the dimensionless mode-dephasing strength $\xim=-\ln\etam$,
\begin{equation}
\QFI^{\rm mode}_{\xim\xim}
=\etam^2\QFI^{\rm mode}_{\etam\etam}
=\frac{4\etam^2P(1-P)}{1-\etam^2}
=\frac{C_{\etam}^2}{1-\etam^2}.
\label{eq:qfi_xi_mode}
\end{equation}
Consequently, estimation of the dimensional rate gives
\begin{equation}
\QFI^{\rm mode}_{\Gamma_{\mathrm m}\Gamma_{\mathrm m}}=L^2\QFI^{\rm mode}_{\xim\xim}
=\frac{4L^2\etam^2P(1-P)}{1-\etam^2}.
\label{eq:qfi_gamma_mode}
\end{equation}
The divergence at $\Gamma_{\mathrm m}\to0^+$ is a nonregular, support-changing boundary effect. Increasing the sample size does not by itself make the per-copy QFI finite; rather, the assumptions underlying the local unbiased Cram\'er--Rao bound cease to be uniform at the boundary. Energy averaging, backgrounds, imperfect control of $L/E$, or a regularized noise model can yield a finite operational problem, while finite-data performance should be assessed with a nonlocal, Bayesian, or other finite-sample bound~\cite{Safranek2017,Jarzyna2015}. A measurement of the flavor-mode populations saturates the $P$ entry of Eq.~\eqref{eq:mode_qfim}, but contains no information about $\chi$ or $\etam$.

\begin{figure}[!htbp]
\centering
\includegraphics[width=0.8\linewidth]{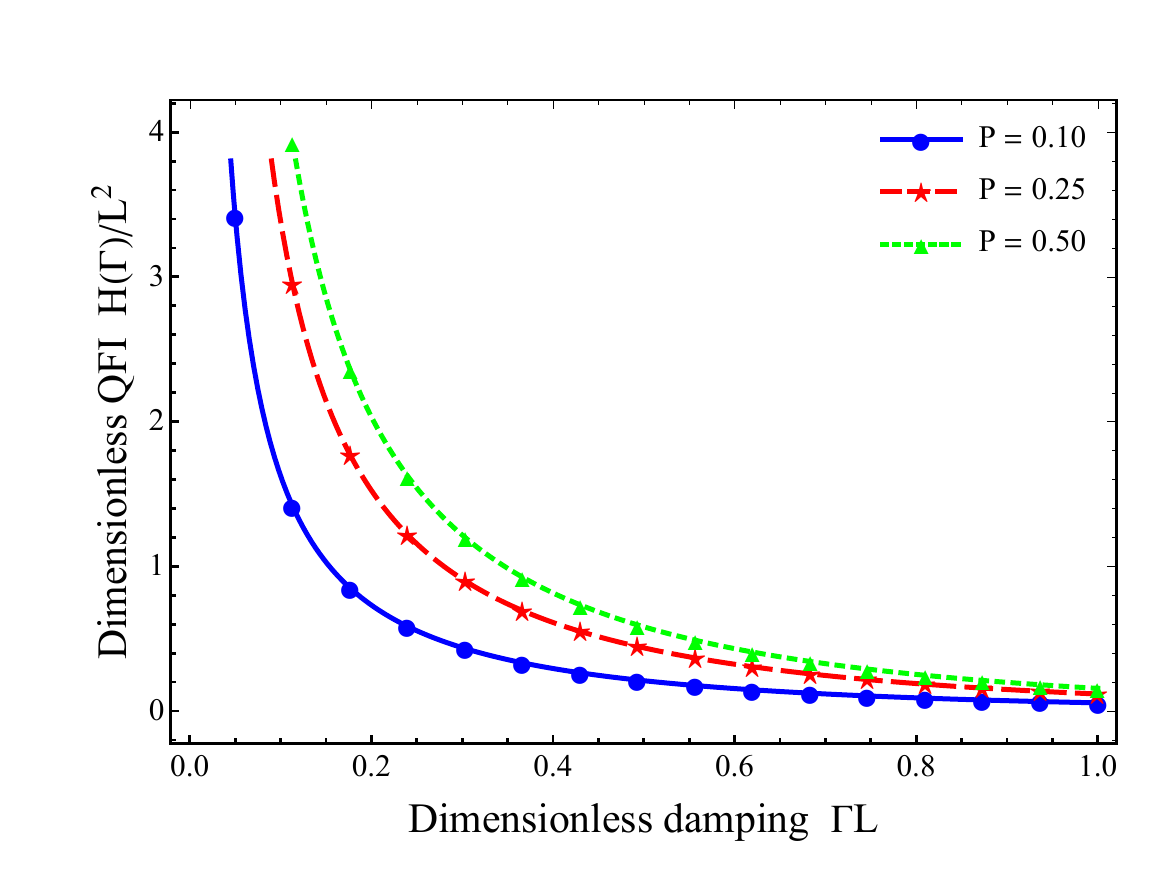}
\caption{Dimensionless QFI for estimating the dephasing strength, $\QFI_{\xim\xim}=\QFI_{\Gamma_{\mathrm m}\Gamma_{\mathrm m}}/L^2$, for fixed transition probability $P$ and $\etam=\exp(-\xim)$.  The idealized QFI is largest near the no-decoherence boundary and decreases as the off-diagonal coherence is lost.}

\label{fig:qfi_decoherence}
\end{figure}
\subsection{Analytic QFI and flavor FI under propagation-basis dephasing}
\label{subsec:qfi-fi-mass-dephasing}
We use the mass-basis dephased state introduced in
Eq.~\eqref{eq:mass_dephasing}, with the damping factor $\etap$ and the
oscillation phase $\Phi$ defined in Eqs.~\eqref{eq:damping_factor} and
\eqref{eq:oscillation_phase}, respectively. This is a Markovian
pure-dephasing channel in the mass basis: the populations are conserved,
whereas the interference term is multiplied by $\etap$. We estimate
$\theta$, $\Delta m^{2}$, and $\Gamma_{\mathrm p}$ separately, with the remaining
parameters held fixed.
For later convenience, define
\begin{equation}
A=\sin^{2}(2\theta),
\qquad
B=1-\etap\cos\Phi,
\qquad
D=1-\frac{A}{2}B.
\label{eq:ABD-definitions}
\end{equation}
The dephased transition probability is then
\begin{equation}
P_{\alpha\rightarrow\beta}^{\mathrm{deph}}
=
\frac{A}{2}B,
\qquad
P_{\alpha\rightarrow\alpha}^{\mathrm{deph}}
=
D.
\label{eq:dephased-probabilities-qfi}
\end{equation}

To compare the state information with a fixed flavor readout, we write
$\rho_f=U(\theta)\rho_mU^\dagger(\theta)$ using the Bloch representation
defined in Eq.~\eqref{eq:bloch_representation}, with
\begin{equation}
\boldsymbol{r}
=
\begin{pmatrix}
\sin(2\theta)\cos(2\theta)\bigl(\etap\cos\Phi-1\bigr)
\\[1mm]
-\etap\sin(2\theta)\sin\Phi
\\[1mm]
\cos^{2}(2\theta)+\etap\sin^{2}(2\theta)\cos\Phi
\end{pmatrix}.
\label{eq:bloch-vector-flavor}
\end{equation}
The mixed-state QFI is evaluated with Eq.~\eqref{eq:qubit_qfi}.
The boundary values $\etap=0$ and $\etap=1$ are obtained by continuity.

For the mixing angle, direct differentiation of
Eq.~\eqref{eq:bloch-vector-flavor} gives
\begin{equation}
\mathcal{H}_{\theta}^{(f)}
=
8\bigl(1-\etap\cos\Phi\bigr)
-
4\sin^{2}(2\theta)
\left(
1-\etap^{2}\cos^{2}\Phi
\right)
.
\label{eq:qfi-theta-dephasing}
\end{equation}
The superscript $(f)$ emphasizes that the derivative is taken after
expressing the state in the fixed flavor basis. By contrast,
differentiation in the fixed mass basis gives the pure-state result
$\mathcal{H}_{\theta}^{(m)}=4$ already obtained in
Eq.~\eqref{eq:qfi_theta_two}. The two expressions differ because the
transformation $U(\theta)$ itself contains the parameter being estimated.
Therefore, the appropriate QFI for comparison with a fixed flavor
measurement is Eq.~\eqref{eq:qfi-theta-dephasing}.

For the mass-squared splitting, the mixing matrix is independent of
$\Delta m^{2}$ and $\partial_{\Delta m^{2}}\Phi=L/(2E)$. The
corresponding QFI is basis invariant and reads
\begin{equation}
\mathcal{H}_{\Delta m^{2}}
=
\etap^{2}\sin^{2}(2\theta)
\left(
\frac{L}{2E}
\right)^{2}
\label{eq:qfi-dm2-dephasing}
\end{equation}

For estimation of the dephasing rate, one uses
$\partial_{\Gamma_{\mathrm p}}\etap=-L\etap$. The propagation-basis QFI becomes
\begin{equation}
\mathcal{H}_{\Gamma_{\mathrm p}}^{\rm prop}
=
\frac{
L^{2}\etap^{2}\sin^{2}(2\theta)
}{
1-\etap^{2}
}.
\label{eq:qfi-gamma-mass-dephasing}
\end{equation}
The divergence as $\Gamma_{\mathrm p}\rightarrow0^{+}$, or equivalently $\etap\rightarrow1^{-}$, is the same nonregular boundary phenomenon. Finite statistics alone do not regularize the per-copy QFI; instead, spectral averaging, backgrounds, systematic uncertainties, prior information, or an explicitly finite-sample estimation criterion change the statistical model and can produce a finite operational bound~\cite{Safranek2017,Jarzyna2015}.

For the two flavor outcomes, the classical FI follows directly from
the binary-outcome expression in Eq.~\eqref{eq:fi_two_outcome}, using
$P=P_{\alpha\rightarrow\beta}^{\mathrm{deph}}$. The required derivatives are
\begin{align}
\partial_{\theta}
P_{\alpha\rightarrow\beta}^{\mathrm{deph}}
&=
\sin(4\theta)\,B,
\label{eq:dtheta-prob-deph}
\\
\partial_{\Delta m^{2}}
P_{\alpha\rightarrow\beta}^{\mathrm{deph}}
&=
\frac{
A\etap L
}{
4E
}
\sin\Phi,
\label{eq:ddm2-prob-deph}
\\
\partial_{\Gamma_{\mathrm p}}
P_{\alpha\rightarrow\beta}^{\mathrm{deph}}
&=
\frac{
AL\etap
}{
2
}
\cos\Phi.
\label{eq:dgamma-prob-deph}
\end{align}
Substitution into Eq.~\eqref{eq:fi_two_outcome} yields
\begin{equation}
\mathcal{F}_{\mathrm{flav}}(\theta)
=
\frac{
\sin^{2}(4\theta)B^{2}
}{
\left(\dfrac{AB}{2}\right)D
}
=
\frac{
8\cos^{2}(2\theta)B
}{
D
}
\label{eq:fi-theta-dephasing}
\end{equation}
\begin{equation}
\mathcal{F}_{\mathrm{flav}}(\Delta m^{2})
=
\frac{
A\etap^{2}
}{
2BD
}
\left(
\frac{L}{2E}
\right)^{2}
\sin^{2}\Phi
,
\label{eq:fi-dm2-dephasing}
\end{equation}
and
\begin{equation}
\mathcal{F}_{\mathrm{flav}}(\Gamma_{\mathrm p})
=
\frac{
AL^{2}\etap^{2}\cos^{2}\Phi
}{
2BD
}
.
\label{eq:fi-gamma-dephasing}
\end{equation}
The corresponding flavor-accessibility ratios are
\begin{equation}
\frac{
\mathcal{F}_{\mathrm{flav}}(\theta)
}{
\mathcal{H}_{\theta}^{(f)}
}
=
\frac{
2(1-A)B
}{
D\left[
2B-A\left(1-\etap^{2}\cos^{2}\Phi\right)
\right]
}
,
\label{eq:efficiency-theta-dephasing}
\end{equation}
\begin{equation}
\frac{
\mathcal{F}_{\mathrm{flav}}(\Delta m^{2})
}{
\mathcal{H}_{\Delta m^{2}}
}
=
\frac{
\sin^{2}\Phi
}{
2BD
}
,
\label{eq:efficiency-dm2-dephasing}
\end{equation}
and
\begin{equation}
\frac{
\mathcal{F}_{\mathrm{flav}}(\Gamma_{\mathrm p})
}{
\mathcal{H}_{\Gamma_{\mathrm p}}^{\rm prop}
}
=
\frac{
(1-\etap^{2})\cos^{2}\Phi
}{
2BD
}
\label{eq:efficiency-gamma-dephasing}
\end{equation}
These expressions separate three effects: the intrinsic information
carried by the dephased neutrino state, the fraction accessible through
flavor projection, and the phase-dependent loss of sensitivity induced
by the mass-basis damping channel.  In particular,
$\mathcal{F}_{\mathrm{flav}}(\Gamma_{\mathrm p})=0$ whenever
$\cos\Phi=0$, because the diagonal flavor probabilities are then locally
independent of $\etap$, even though the state-level QFI
$\mathcal{H}_{\Gamma_{\mathrm p}}^{\rm prop}$ remains nonzero.

If the phenomenological rate is written as
$\Gamma_{\mathrm p}(E)=\Gamma_{0}(E/E_{0})^{n}$ and $\Gamma_{0}$ is the estimated
parameter, Eqs.~\eqref{eq:qfi-gamma-mass-dephasing} and
\eqref{eq:fi-gamma-dephasing} remain valid after the replacement
\begin{equation}
L
\longrightarrow
L\left(\frac{E}{E_{0}}\right)^{n}
\end{equation}
in every derivative with respect to $\Gamma_{0}$.
\FloatBarrier
\section{Three-flavor formulation, numerical protocol, and benchmark results}
\label{sec:results}

All numerical results in this section are monochromatic, state-level
information-geometric diagnostics. They do not include fluxes, cross sections,
energy migration, detector response, backgrounds, systematic uncertainties,
or nuisance-parameter marginalization and therefore are not event-level
sensitivity forecasts.

\subsection{Three-flavor mixed-state QFIM in constant-density matter}
\label{subsec:threeflavor_open_qfim}

To connect the analytic two-flavor results to long-baseline precision measurements, we use a constant-density three-flavor benchmark inspired by the published design parameters of DUNE~\cite{DUNECDR}, T2HK~\cite{T2HKDesign}, and ESSnuSB~\cite{ESSnuSB}.  In the flavor basis the effective mass-squared operator is
\begin{align}
\mathsf{M}^{2}_{f}(E)&=U\,\diag(0,\Delta m_{21}^{2},\Delta m_{31}^{2})U^{\dagger}
+\diag(a,0,0),\\
a&=2\sqrt{2}G_{F}N_{e}E.
\label{eq:matter_mass_operator}
\end{align}
and $\mathsf{M}^{2}_{f}=V\,\diag(\widetilde m_1^2,\widetilde m_2^2,\widetilde m_3^2)V^{\dagger}$.  For an initially produced flavor state $\rho_f(0)=\proj{\nu_\alpha}$, we define $\widetilde\rho(0)=V^{\dagger}\rho_f(0)V$ and evolve the effective propagation-basis elements as
\begin{equation}
\widetilde\rho_{ij}(L)=\widetilde\rho_{ij}(0)
\exp\!\left[-\ii\frac{(\widetilde m_i^2-\widetilde m_j^2)L}{2E}\right]
\exp[-\Gamma_{ij}(E)L],
\label{eq:threeflavor_dephased_state}
\end{equation}
Here $\Gamma_{ii}=0$, and the benchmark uses the common off-diagonal rate $\Gamma_{ij}=\Gamma_0(E/E_0)^n$ for $i\neq j$. Equation~\eqref{eq:threeflavor_dephased_state}, rather than Eq.~\eqref{eq:gellmann_dissipator}, is the operational definition of the implemented numerical channel. At each parameter point, $\mathsf M_f^2(\boldsymbol\lambda)$ is diagonalized and $V(\boldsymbol\lambda)$ is used to define the propagation basis; recomputing $V$ under differentiation therefore changes both the Hamiltonian eigenvectors and the basis in which dephasing is imposed. This is a physical model choice, not a passive representation change. A realistic analysis should allow independent $(\Gamma_{21},\Gamma_{31},\Gamma_{32})$ subject to complete-positivity constraints. The detector-basis state is $\rho_f(L)=V\widetilde\rho(L)V^{\dagger}$, and antineutrino propagation follows from $U\to U^{*}$ and $a\to-a$.

We use the dimensionless parameter vector
\begin{align}
\boldsymbol{\lambda}&=(\theta_{23},\delta_{\rm CP},\xi_{31},g),\\
\xi_{31}&=\ln\!\frac{|\Delta m_{31}^{2}|}{\Delta m_{31,\rm ref}^{2}},\\
g&=\log_{10}\!\frac{\Gamma_0}{10^{-23}\,\mathrm{GeV}}.
\label{eq:dimensionless_parameter_vector}
\end{align}
The logarithmic coordinates avoid unit-dependent comparisons and remove the formal $\Gamma_0\to0$ divergence that appears when the rate itself is used as a coordinate.

Following the general mixed-state quantum-estimation formulation of Ref.~\cite{ParisReview} and its application to neutrino oscillation parameters in Ref.~\cite{FrugiueleParis2026}, we evaluate the QFIM of $\rho=\sum_m p_m\ket{m}\bra{m}$ as
\begin{equation}
\mathcal{H}_{ab}=2\sum_{m,n}\frac{\Re\!\left[\bra{m}\partial_a\rho\ket{n}
\bra{n}\partial_b\rho\ket{m}\right]}{p_m+p_n},
\qquad p_m+p_n>0,
\label{eq:mixed_qfim_spectral}
\end{equation}
whereas ideal flavor projection gives
\begin{equation}
\mathcal{F}^{\rm flav}_{ab}=\sum_{\beta=e,\mu,\tau}
\frac{\partial_a P_{\alpha\beta}\,\partial_b P_{\alpha\beta}}{P_{\alpha\beta}}.
\label{eq:threeflavor_flavor_fim}
\end{equation}
For a single neutrino energy and ideal flavor readout, the three probabilities obey $\sum_{\beta}P_{\alpha\beta}=1$. If $J_{\beta a}=\partial_aP_{\alpha\beta}$ denotes the probability Jacobian, only two rows are independent and
$\CFI^{\rm flav}=J^{\mathsf T}\operatorname{diag}(1/P_{\alpha\beta})J$ has rank at most two. The $4\times4$ flavor FIM for $(\theta_{23},\delta_{\rm CP},\xi_{31},g)$ is therefore singular at every monochromatic point and cannot support joint unbiased identification of all four coordinates. The diagonal ratios reported below are conditional, fixed-coordinate diagnostics. A nonsingular joint FIM requires additional independent settings, such as energy bins, neutrino and antineutrino modes, baselines, or oscillation channels.\\
For simultaneous estimation, the SLD Cram\'er--Rao matrix bound is $\mathrm{Cov}(\widehat{\boldsymbol\lambda})\succeq\mathcal{H}^{-1}/N$.  Its joint attainability must be checked.  We therefore calculate
\begin{equation}
\mathcal{D}_{ab}=\frac{1}{2\ii}\Tr\!\left(\rho[L_a,L_b]\right),
\qquad
\mathfrak{c}_{ab}=\frac{|\mathcal{D}_{ab}|}{\sqrt{\mathcal{H}_{aa}\mathcal{H}_{bb}}},
\label{eq:sld_compatibility}
\end{equation}
where $L_a$ is the SLD. Nonzero $\mathcal{D}_{ab}$ is a warning that the corresponding single-parameter quantum limits are not generally saturable by one common measurement~\cite{Ragy2016,Candeloro,Asjad,BelliardoGiovannetti2021}. The pairwise quantity $\mathfrak c_{ab}$ is a useful diagnostic, but it does not quantify the full gap between the SLD matrix bound and the attainable multiparameter precision. The Holevo Cram\'er--Rao bound, or an explicit optimization over a physically allowed compatible measurement family, is required for an operational joint limit~\cite{Albarelli2019,Xia2023}.

The reported precision penalty treats the dephasing rate as fixed and known.
Using the standard-parameter submatrix $\QFI_{\rm std}$, we define
\begin{equation}
\Pi^{\rm known}_a(\Gamma_0)=
\sqrt{\frac{[\QFI_{\rm std}^{-1}(\Gamma_0)]_{aa}}
{[\QFI_{\rm std}^{-1}(\Gamma_{\rm ref})]_{aa}}},
\label{eq:precision_penalty}
\end{equation}
with $\Gamma_{\rm ref}=10^{-25}\,\mathrm{GeV}$. Values above unity quantify a
degradation of the SLD quantum Cram\'er--Rao matrix lower bound. They are not
generally jointly attainable when the SLDs are incompatible. A separate
four-parameter inversion in which $g$ is estimated and marginalized is not
reported here and is identified below as a required extension.

\subsection{Numerical inputs and reproducibility protocol}
\label{subsec:numerical_protocol}
All angular parameters are evaluated in radians. The vacuum phase and matter
term are implemented as
\begin{align}
\Phi_{ij}&=2.53386535884
\left(\frac{\Delta m^2_{ij}}{\mathrm{eV}^2}\right)
\left(\frac{L}{\mathrm{km}}\right)
\left(\frac{\mathrm{GeV}}{E}\right),
\label{eq:numerical_phase}\\
\frac{a}{\mathrm{eV}^2}&=1.52\times10^{-4}Y_e
\left(\frac{\rho}{\mathrm{g\,cm^{-3}}}\right)
\left(\frac{E}{\mathrm{GeV}}\right),
\label{eq:numerical_matter}
\end{align}
with $Y_e=0.5$ for the crust benchmarks. The damping exponent uses
$1\,\mathrm{km}=5.06773071616\times10^{18}\,\mathrm{GeV}^{-1}$. The numerical
inputs used for the three-flavor panels are listed in
Table~\ref{tab:threeflavor_inputs}.

\begin{table*}
\caption{Three-flavor single-energy benchmark inputs used for the QFIM,
flavor-FIM, and SLD-compatibility calculations. The oscillation parameters entering these calculations are the fixed legacy reference set specified below and are kept unchanged throughout all numerical figures and tables. The ESSnuSB-inspired second-maximum point is treated as a vacuum benchmark in the present analysis..}
\label{tab:threeflavor_inputs}
\centering
\begin{tabular}{lcccccc}
\toprule
Benchmark & $L$ (km) & $E$ (GeV) & $\rho$ (g cm$^{-3}$) & Initial state & $n$ & $\Gamma_0$ (GeV)\\
\midrule
DUNE-like & 1300 & 2.50 & 2.85 & $\nu_\mu$ & 0 & $10^{-23}$\\
T2HK-like & 295 & 0.60 & 2.60 & $\nu_\mu$ & 0 & $10^{-23}$\\
ESSnuSB-inspired vacuum & 360 & 0.25 & 0 & $\nu_\mu$ & 0 & $10^{-23}$\\
\bottomrule
\end{tabular}
\end{table*}

The common oscillation inputs are $\theta_{12}=33.44^\circ$, $\theta_{13}=8.57^\circ$, $\theta_{23}=49.20^\circ$, $\delta_{\rm CP}=195^\circ$, $\Delta m^2_{21}=7.42\times10^{-5}\,\mathrm{eV}^2$, and $\Delta m^2_{31}=2.517\times10^{-3}\,\mathrm{eV}^2$ in normal ordering. These values are retained here as a legacy benchmark consistent with the 2020 NuFIT analysis~\cite{NuFIT2020}; they should not be attributed to the distinct NuFIT 6.0 update~\cite{NuFIT60}.

Derivatives are evaluated with centered finite differences in the
dimensionless coordinates $(\theta_{23},\delta_{\rm CP},\xi_{31},g)$.
The angular steps are $h_{\theta}=h_{\delta}=10^{-5}\,\mathrm{rad}$,
while $h_{\xi}=h_g=10^{-5}$. For the logarithmic dephasing coordinate,
$\partial_g\rho=(\ln 10)\Gamma_0\partial_{\Gamma_0}\rho$. Each
derivative is recomputed with $h/2$ and $h/4$. Eigenvalue pairs with
$p_m+p_n\leq10^{-12}$ are omitted from the spectral QFIM.

To quantify finite-difference convergence, we distinguish derivative-level
and information-matrix stability. For
$a\in\{\theta_{23},\delta_{\rm CP},\xi_{31},g\}$, define
\begin{equation}
\epsilon_{\partial\rho}
=
\max_{\substack{a\\ s\in\{h,h/2\}}}
\frac{
\left\|(\partial_a\rho)_{s/2}-(\partial_a\rho)_s\right\|_{F}
}{
\max\!\left[
\left\|(\partial_a\rho)_{s/2}\right\|_{F},\,10^{-30}
\right]
},
\label{eq:derivative_convergence}
\end{equation}
where $\|\cdot\|_{F}$ denotes the Frobenius norm. The corresponding
matrix-level indicator is
\begin{equation}
\epsilon_{\rm mat}
=
\max_{\substack{X\in\{\mathcal{H},\,\mathcal{F}^{\rm flav}\}\\
s\in\{h,h/2\}}}
\frac{
\left\|X_{s/2}-X_s\right\|_{F}
}{
\max\!\left[
\left\|X_{s/2}\right\|_{F},\,10^{-30}
\right]
}.
\label{eq:matrix_convergence}
\end{equation}
Thus, $\epsilon_{\partial\rho}$ measures the largest relative change in
the density-matrix derivatives, whereas $\epsilon_{\rm mat}$ measures the
largest relative change in the resulting QFIM or flavor FIM under the
successive refinements $h\rightarrow h/2\rightarrow h/4$.

At every benchmark point, the implementation additionally checks
normalization, Hermiticity, positivity of $\rho$, positive
semidefiniteness of $\mathcal{H}$, and positive semidefiniteness of
$\mathcal{H}-\mathcal{F}^{\rm flav}$ for the same statistical encoding.
The numerical rank of the monochromatic flavor FIM is obtained from its
singular-value decomposition using a relative threshold
$s_i/s_1>10^{-10}$. The corresponding validation diagnostics are reported
in Table~\ref{tab:numerical_validation}.

\begin{table*}
\centering
\caption{Numerical validation diagnostics for the three-flavor
benchmark points at $\Gamma_0=10^{-23}\,\mathrm{GeV}$. The first two
columns quantify normalization and Hermiticity errors. The quantities
$\lambda_{\min}(\rho)$, $\lambda_{\min}(\mathcal{H})$, and
$\lambda_{\min}(\mathcal{H}-\mathcal{F}^{\rm flav})$ test positivity
of the propagated state, the QFIM, and the QFI--FI difference,
respectively. The indicators $\epsilon_{\partial\rho}$ and
$\epsilon_{\rm mat}$ are defined in
Eqs.~\eqref{eq:derivative_convergence} and
\eqref{eq:matrix_convergence}. The last column reports the numerical
rank of the monochromatic flavor FIM. All minimum eigenvalues remain
non-negative to the reported precision.}
\label{tab:numerical_validation}
\resizebox{\textwidth}{!}{%
\begin{tabular}{lcccccccc}
\toprule
Benchmark
& $|\Tr\rho-1|$
& $\|\rho-\rho^\dagger\|_{F}$
& $\lambda_{\min}(\rho)$
& $\lambda_{\min}(\mathcal{H})$
& $\lambda_{\min}(\mathcal{H}-\mathcal{F}^{\rm flav})$
& $\epsilon_{\partial\rho}$
& $\epsilon_{\rm mat}$
& $\operatorname{rank}(\mathcal{F}^{\rm flav})$ \\
\midrule
DUNE-like
& $6.7\times10^{-16}$
& $0$
& $3.16\times10^{-3}$
& $1.71\times10^{-1}$
& $2.04\times10^{-2}$
& $4.66\times10^{-10}$
& $3.18\times10^{-10}$
& $2$ \\

T2HK-like
& $8.9\times10^{-16}$
& $0$
& $2.06\times10^{-3}$
& $4.40\times10^{-2}$
& $1.03\times10^{-2}$
& $1.03\times10^{-9}$
& $2.97\times10^{-10}$
& $2$ \\

ESSnuSB-inspired vacuum
& $4.4\times10^{-16}$
& $0$
& $1.82\times10^{-3}$
& $5.19\times10^{-2}$
& $1.04\times10^{-2}$
& $1.09\times10^{-9}$
& $9.26\times10^{-10}$
& $2$ \\
\bottomrule
\end{tabular}%
}

\end{table*}

The validation results demonstrate that the numerical implementation
preserves the density-matrix constraints to machine precision. In
particular, the trace and Hermiticity errors are negligible, while the
non-negative minimum eigenvalues of $\rho$, $\mathcal{H}$, and
$\mathcal{H}-\mathcal{F}^{\rm flav}$ provide independent numerical
checks of state positivity, QFIM positive-semidefiniteness, and the
QFI--FI information inequality for the same encoding. The values of
$\epsilon_{\partial\rho}$ and $\epsilon_{\rm mat}$, which are of order
$10^{-9}$ or smaller, further show that both the density-matrix
derivatives and the resulting information matrices are stable under
finite-difference refinement.

As an additional structural check, the numerical flavor FIM has rank
$2$ at each of the DUNE-like, T2HK-like, and ESSnuSB-inspired vacuum
benchmark points. This agrees with the analytical bound
$\operatorname{rank}(\mathcal{F}^{\rm flav})\leq2$ that follows from
$\sum_{\beta}P_{\alpha\beta}=1$ for a single-energy three-outcome flavor
measurement. Consequently, the four diagonal flavor-accessibility
ratios reported below must be interpreted as conditional
fixed-coordinate diagnostics rather than as a jointly invertible
four-parameter measurement model.

\begin{figure*}
		{{\begin{minipage}[t]{.33\linewidth}
					\centering
					(a)\par\vspace{1mm}
\includegraphics[width=\linewidth]{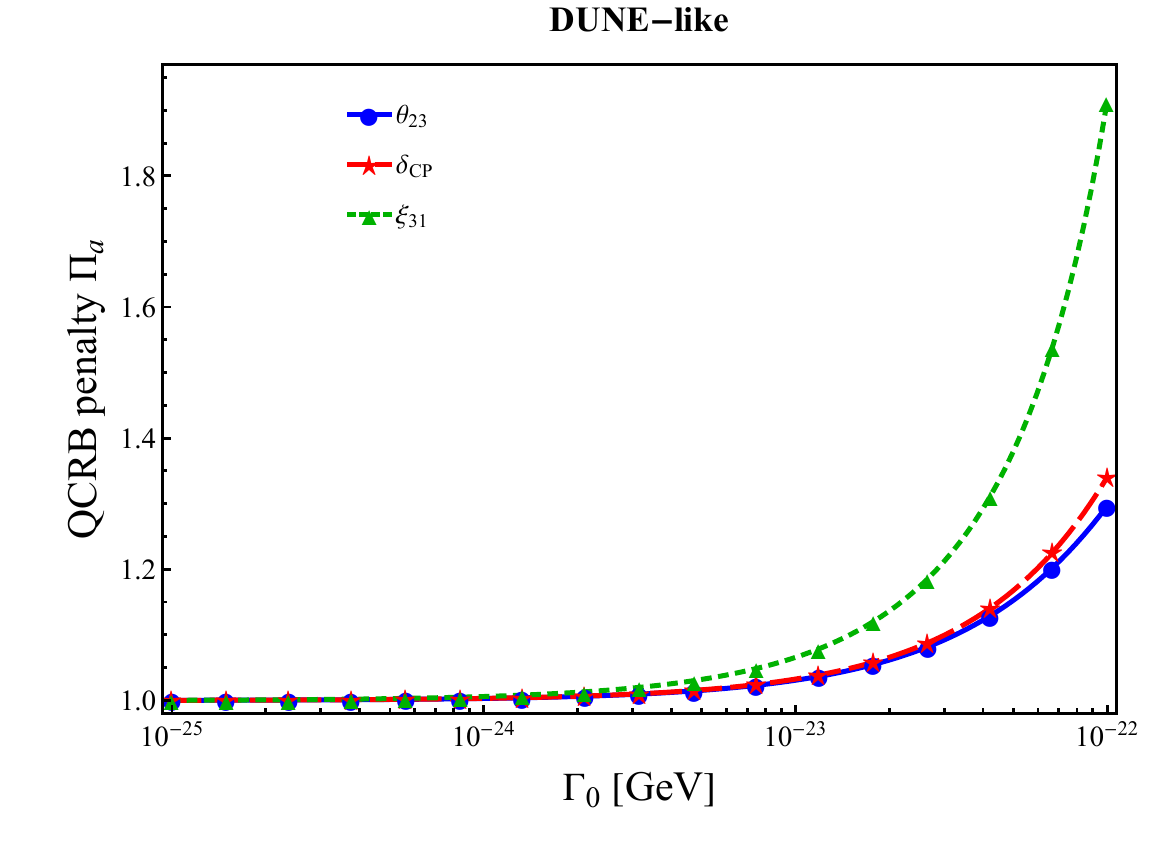}
				\end{minipage}\hfill
				\begin{minipage}[t]{.33\linewidth}
					\centering
					(b)\par\vspace{1mm}
\includegraphics[width=\linewidth]{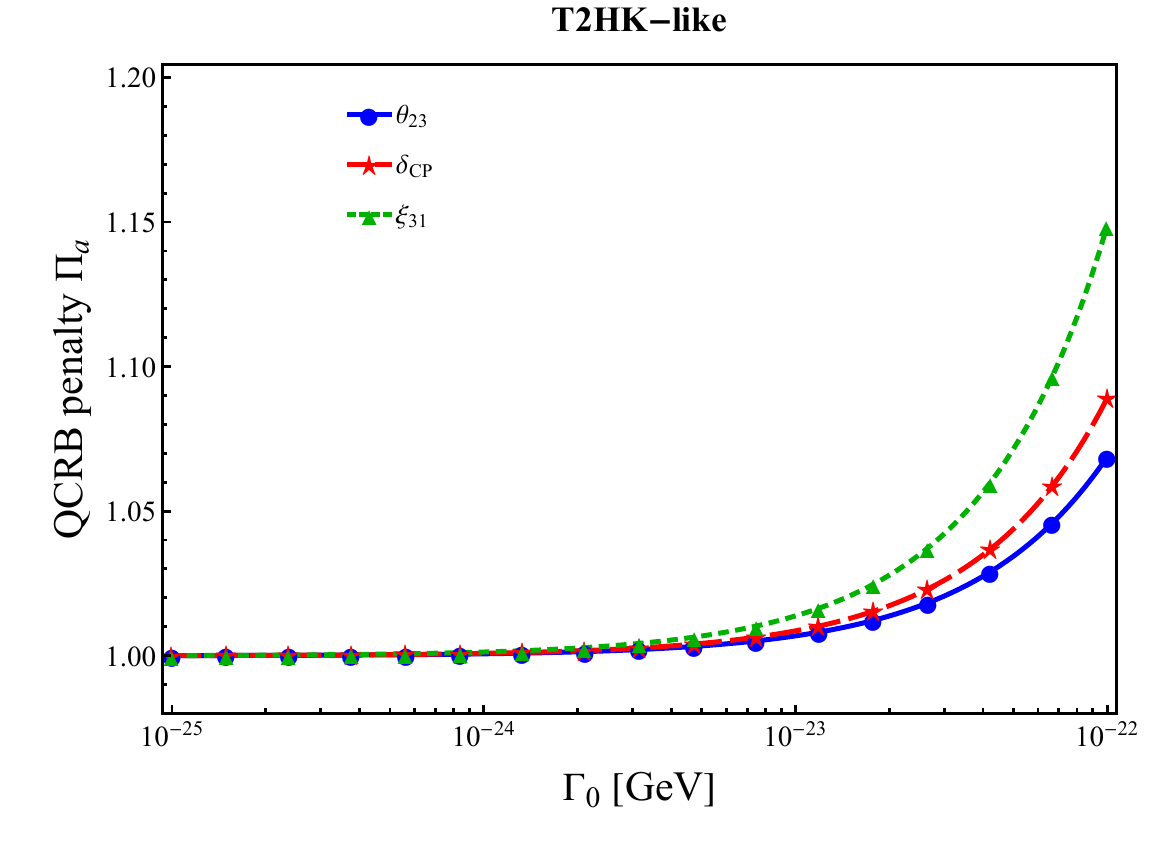}
		\end{minipage}}
	\begin{minipage}[t]{.33\linewidth}
		\centering
		(c)\par\vspace{1mm}
\includegraphics[width=\linewidth]{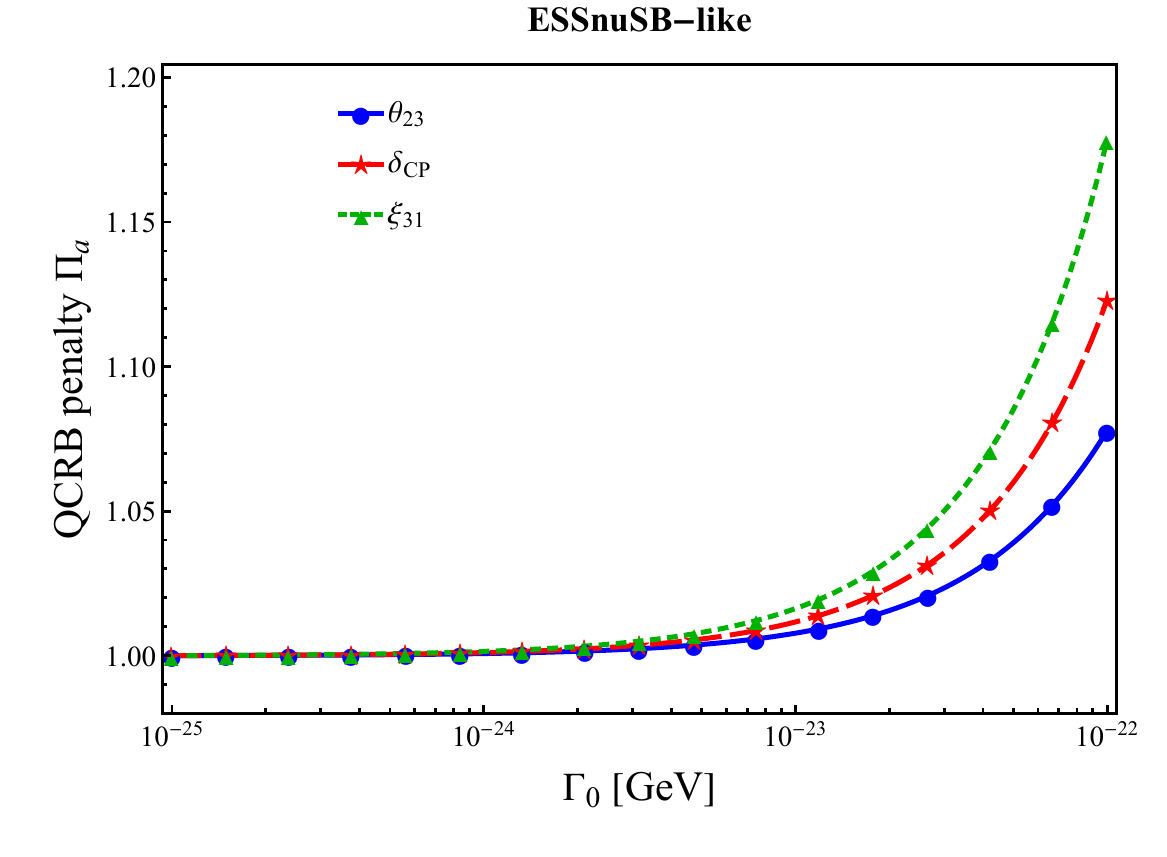}
\end{minipage}}
\caption{Known-rate SLD-QCRB penalty $\Pi^{\rm known}_a$ for $a\in\{\theta_{23},\delta_{\rm CP},\xi_{31}\}$ as a function of the common dephasing rate. In this single-energy constant-density benchmark, DUNE-like propagation develops the largest high-rate penalty, while the T2HK-like constant-density configuration and the
ESSnuSB-inspired vacuum configuration are less affected over the
displayed range.}
\label{fig:qcrb_penalty}
\end{figure*}

 At $\Gamma_{0}=10^{-23}\,\mathrm{GeV}$, the diagonal fixed-coordinate flavor-accessibility ratios $\mathcal{F}^{\rm flav}_{aa}/\mathcal{H}_{aa}$ for $(\theta_{23},\delta_{\rm CP},\xi_{31},g)$ are approximately $(0.30,0.010,0.047,0.61)$ for the DUNE-like point, $(0.66,0.013,0.081,0.22)$ for the T2HK-like point, and $(0.65,0.105,0.095,0.21)$ for the ESSnuSB-inspired second-maximum vacuum point. These numbers are deliberately reported as benchmark diagnostics rather than universal experimental efficiencies: spectral convolution, antineutrino data, detector effects, priors, and nuisance parameters can change them substantially.

Figure~\ref{fig:qcrb_penalty} shows the known-rate SLD-QCRB penalty in Eq.~\eqref{eq:precision_penalty} for $a\in\{\theta_{23},\delta_{\rm CP},\xi_{31}\}$. A value $\Pi_a=1$ denotes no degradation relative to the nearly coherent reference, whereas $\Pi_a>1$ means that the SLD matrix lower bound worsens because of decoherence and multiparameter correlations; joint attainability is not implied. At the selected benchmark points, the DUNE-like configuration develops the largest high-rate penalty, consistent with its larger accumulated factor $\Gamma_0L$. The comparison, however, changes baseline, energy, oscillation phase, and matter prescription simultaneously, so it does not isolate baseline as the sole cause. The T2HK-like and ESSnuSB-inspired points remain closer to the coherent reference over the displayed range.

\begin{figure*}
\centering


\begin{minipage}[t]{0.3\textwidth}
\centering
(a)\par\vspace{1mm}
\includegraphics[width=\linewidth]
{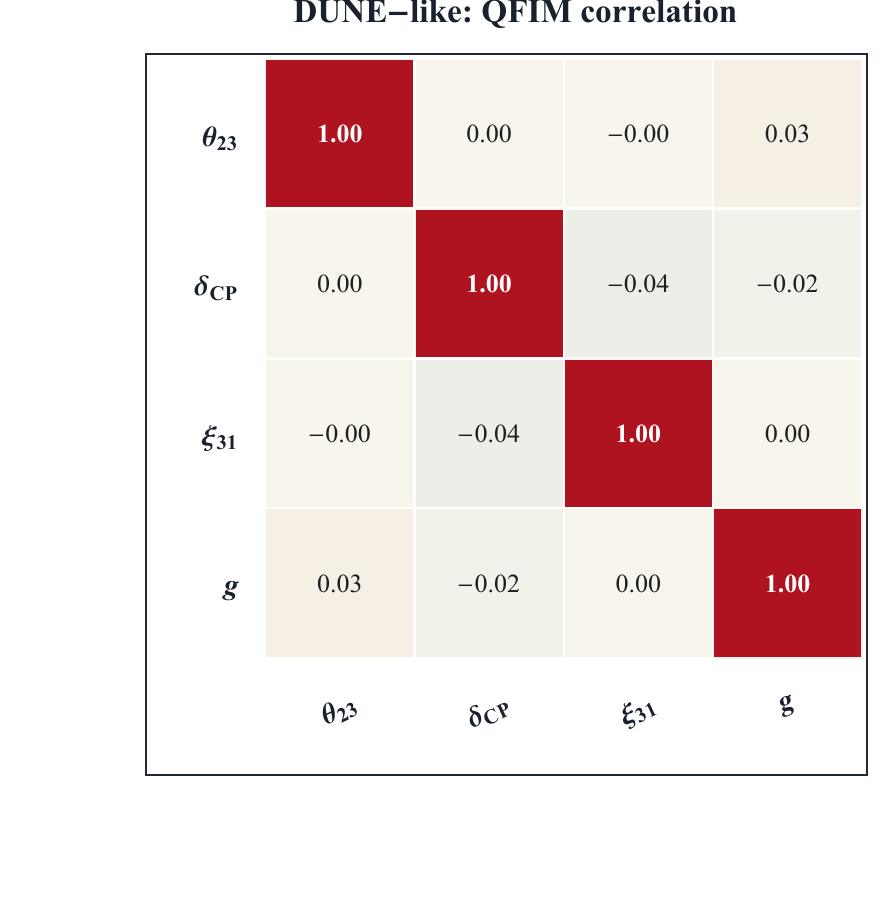}
\end{minipage}
\hfill
\begin{minipage}[t]{0.3\textwidth}
\centering
(b)\par\vspace{1mm}
\includegraphics[width=\linewidth]
{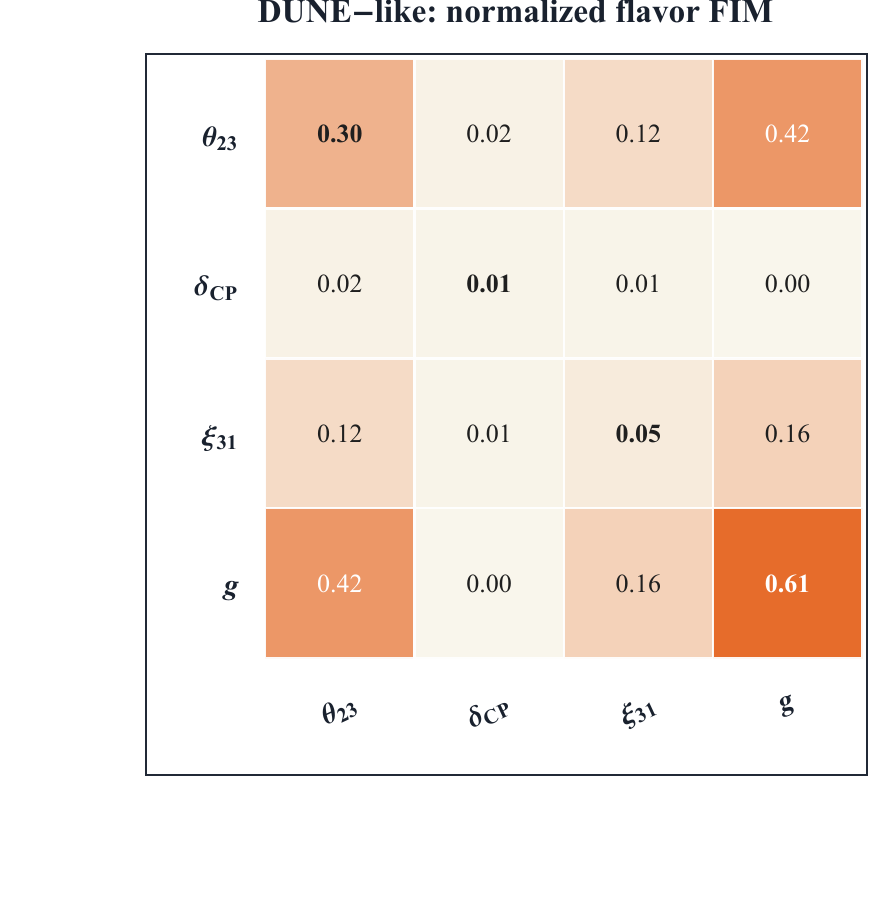}
\end{minipage}
\hfill
\begin{minipage}[t]{0.3\textwidth}
\centering
(c)\par\vspace{1mm}
\includegraphics[width=\linewidth]
{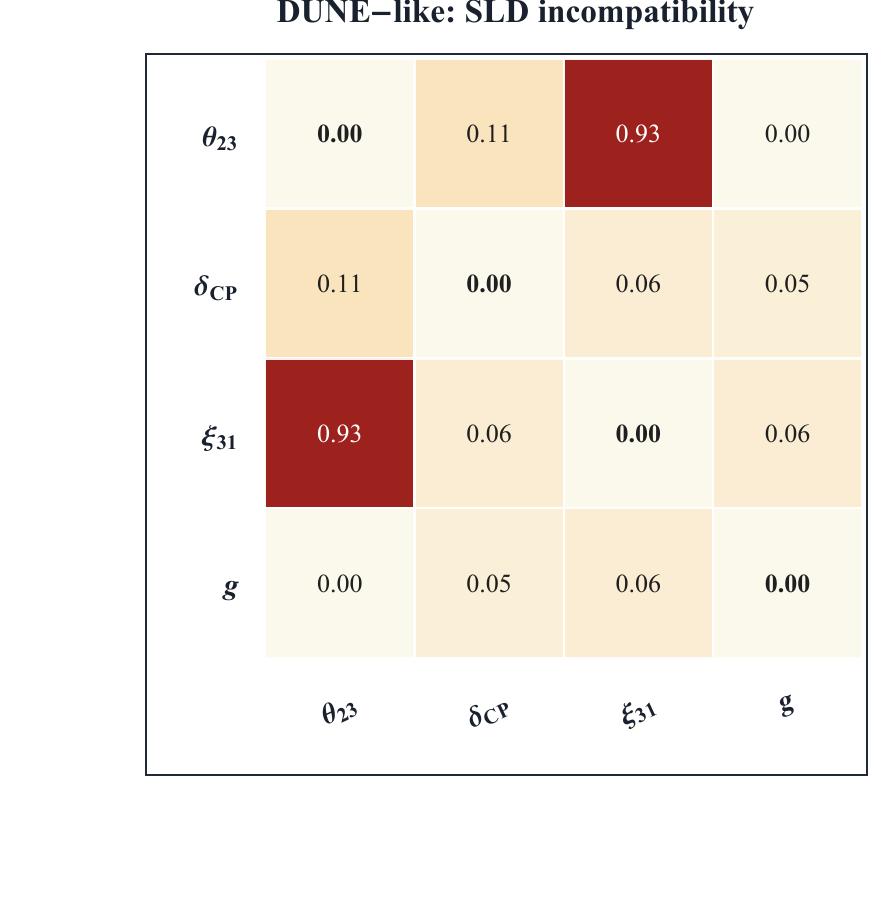}
\end{minipage}

\vspace{1mm}


\begin{minipage}[t]{0.3\textwidth}
\centering
(d)\par\vspace{1mm}
\includegraphics[width=\linewidth]
{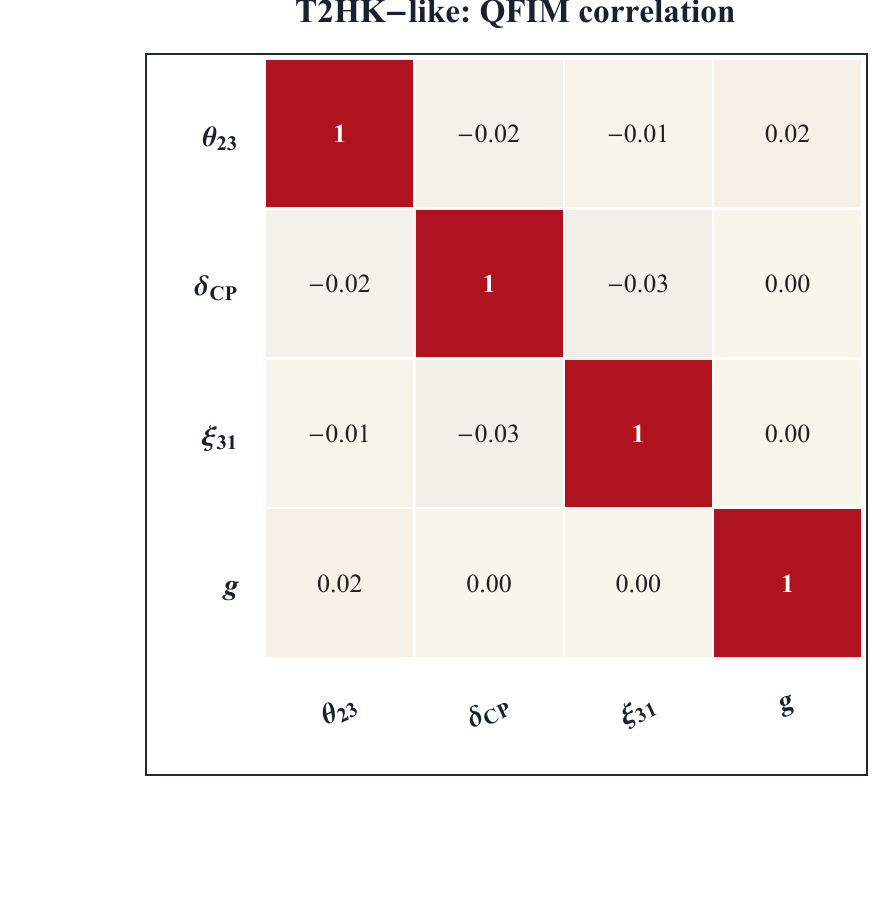}
\end{minipage}
\hfill
\begin{minipage}[t]{0.3\textwidth}
\centering
(e)\par\vspace{1mm}
\includegraphics[width=\linewidth]
{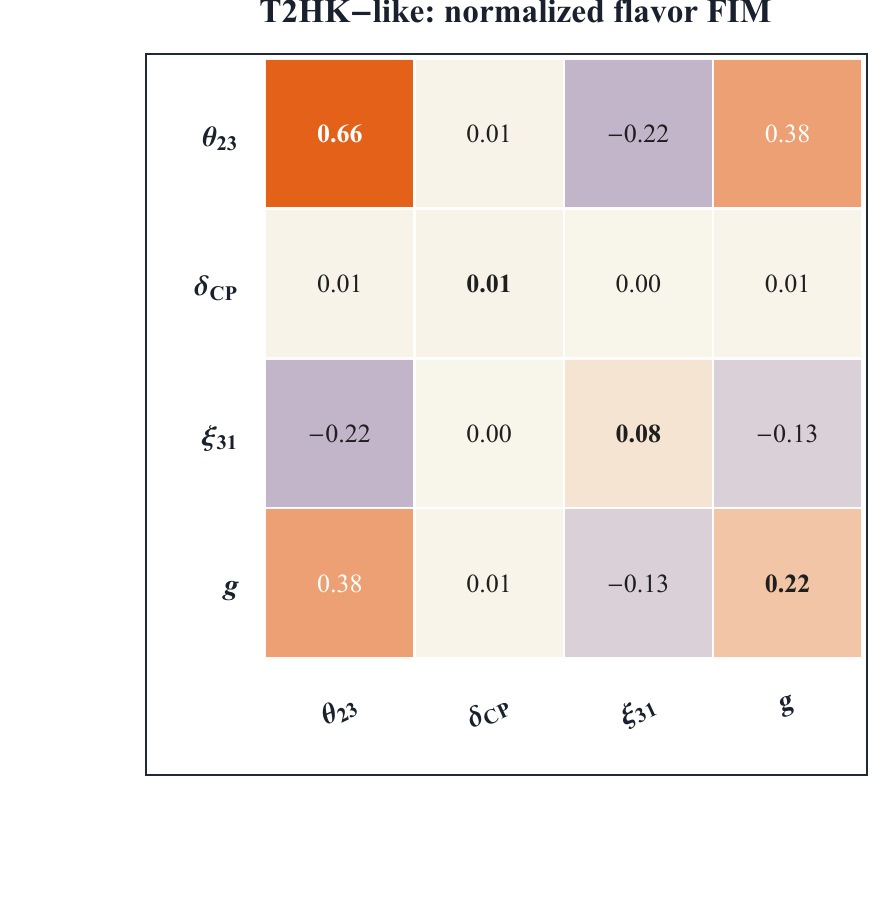}
\end{minipage}
\hfill
\begin{minipage}[t]{0.3\textwidth}
\centering
(f)\par\vspace{1mm}
\includegraphics[width=\linewidth]
{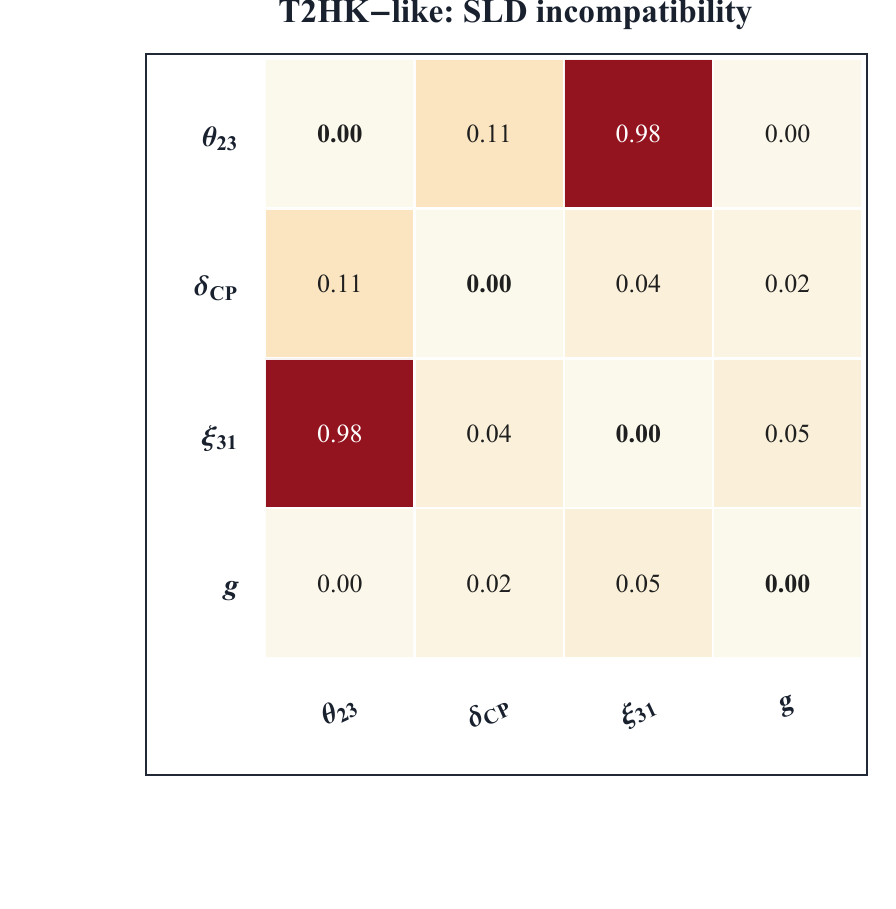}
\end{minipage}

\vspace{1mm}


\begin{minipage}[t]{0.3\textwidth}
\centering
(g)\par\vspace{1mm}
\includegraphics[width=\linewidth]
{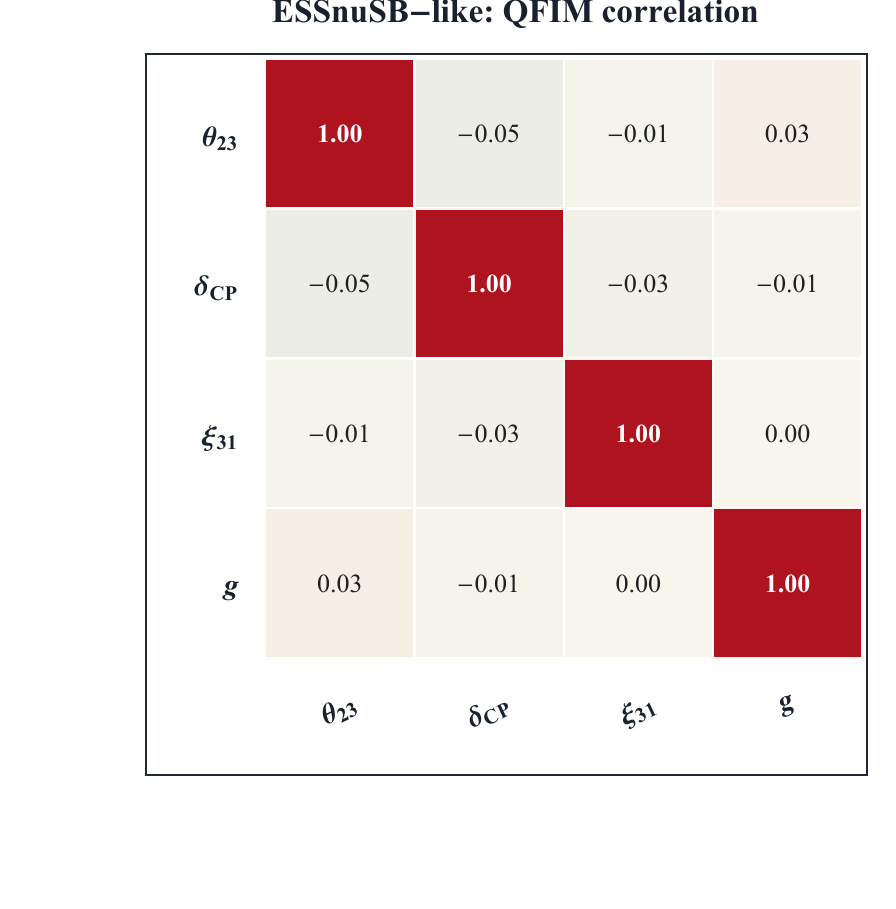}
\end{minipage}
\hfill
\begin{minipage}[t]{0.3\textwidth}
\centering
(h)\par\vspace{1mm}
\includegraphics[width=\linewidth]
{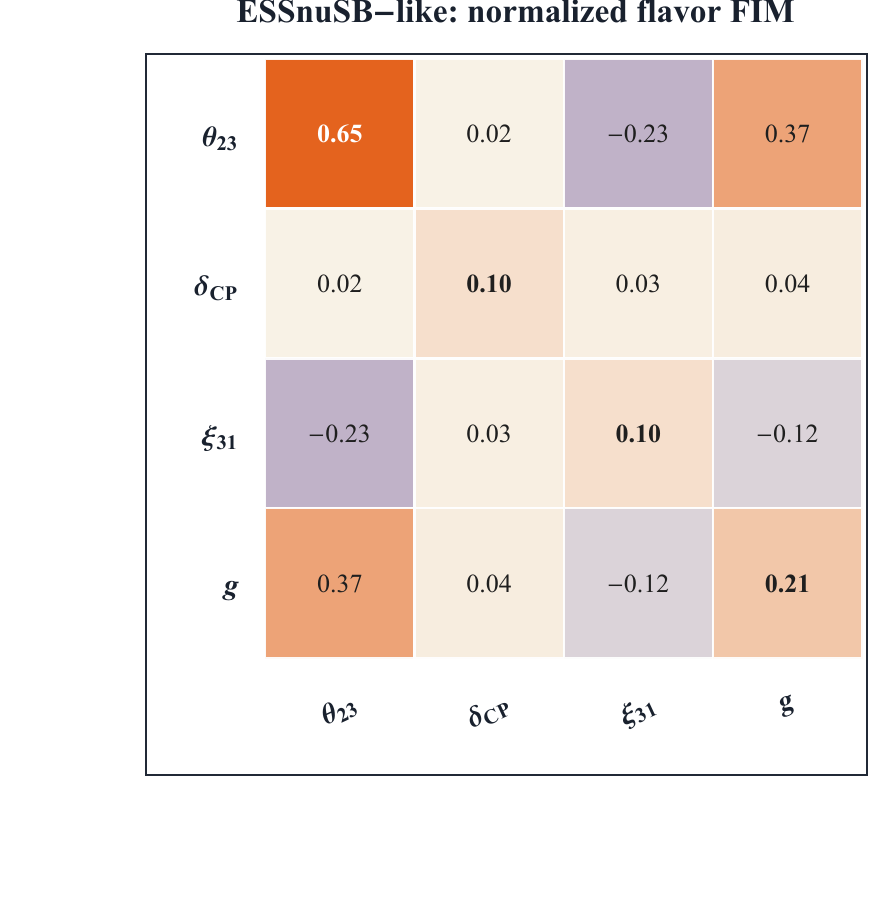}
\end{minipage}
\hfill
\begin{minipage}[t]{0.3\textwidth}
\centering
(i)\par\vspace{1mm}
\includegraphics[width=\linewidth]
{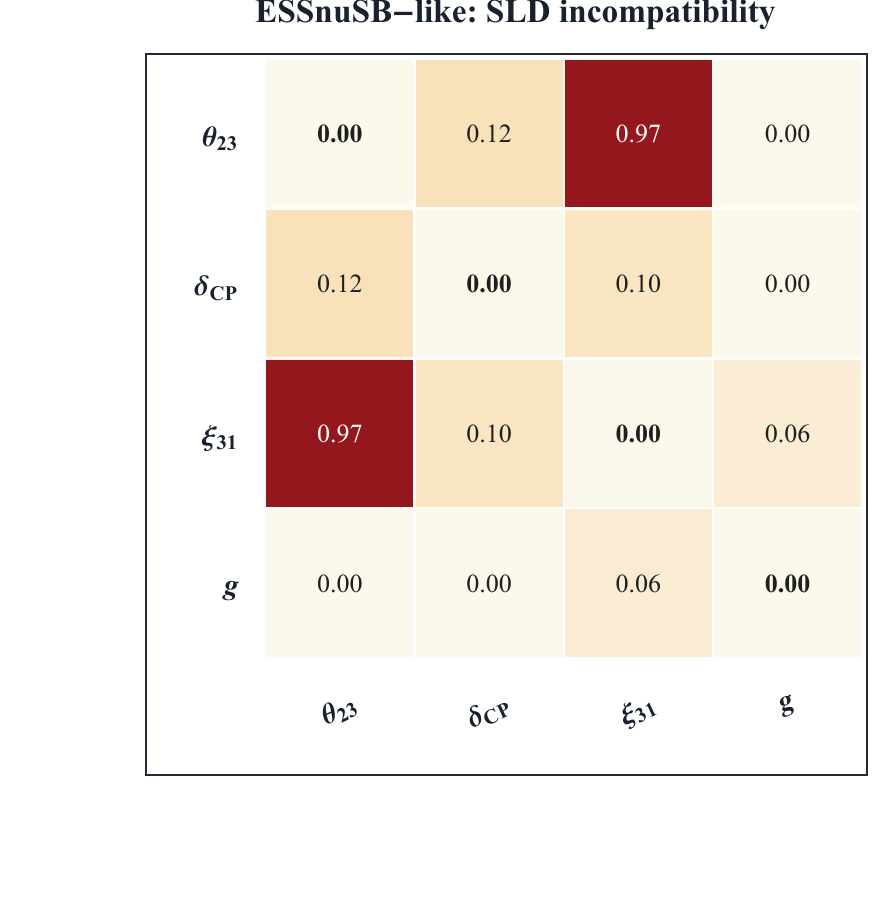}
\end{minipage}

\caption{Three-flavor constant-density benchmarks at
$\Gamma_{0}=10^{-23}\,\mathrm{GeV}$ for the DUNE-like
(top row), T2HK-like (middle row), and ESSnuSB-inspired vacuum
(bottom row) configurations. The left column shows the
normalized QFIM correlations
$\mathcal{H}_{ab}/\sqrt{\mathcal{H}_{aa}\mathcal{H}_{bb}}$;
the middle column shows the flavor FIM normalized by
$\sqrt{\mathcal{H}_{aa}\mathcal{H}_{bb}}$; and the right
column shows the normalized SLD incompatibility
$c_{ab}$. The comparison demonstrates that QFIM
correlation, flavor accessibility, and joint-estimation
compatibility depend differently on baseline, energy,
oscillation phase, and matter effects. In all three
configurations, the strong
$\theta_{23}$--$\xi_{31}$ SLD incompatibility shows why
diagonal QFI curves alone are insufficient for simultaneous
estimation.}
\label{fig:threeflavor_qfim_decoherence}
\end{figure*}

\begin{table*}
\caption{Diagonal flavor-accessibility ratios and the dominant normalized SLD
incompatibility at $\Gamma_0=10^{-23}\,\mathrm{GeV}$. The entries are
monochromatic state-level diagnostics.}
\label{tab:threeflavor_summary}
\centering
\begin{tabular}{lccccc}
\toprule
Benchmark & $\CFI_{\theta_{23}\theta_{23}}/\QFI_{\theta_{23}\theta_{23}}$
& $\CFI_{\delta\delta}/\QFI_{\delta\delta}$
& $\CFI_{\xi\xi}/\QFI_{\xi\xi}$
& $\CFI_{gg}/\QFI_{gg}$
& $\mathfrak c_{\theta_{23},\xi_{31}}$\\
\midrule
DUNE-like & 0.30 & 0.010 & 0.047 & 0.61 & 0.93\\
T2HK-like & 0.66 & 0.013 & 0.081 & 0.22 & 0.98\\
ESSnuSB-inspired vacuum & 0.65 & 0.105 & 0.095 & 0.21 & 0.97\\
\bottomrule
\end{tabular}
\end{table*}

Table~\ref{tab:threeflavor_summary} makes two conclusions quantitative. First,
ideal flavor projection extracts a diagonal fixed-coordinate ratio of about one percent on $\delta_{\rm CP}$ at the nominal DUNE- and T2HK-like points,
whereas the ESSnuSB-inspired second-maximum vacuum point reaches about ten percent.
Second, the large $\theta_{23}$--$\xi_{31}$ incompatibility shows that modest
real QFIM correlations do not imply compatible optimal measurements.

\subsection{Resource survival versus metrological survival}
\label{subsec:resource_metrology}

To compare resource survival with a finite and well-defined metrological quantity, we normalize the correlation measures and the mode-state QFI for the coherent phase $\chi$:
\begin{equation}
\widetilde{\mathcal{M}}_k(\xim)
=
\frac{\mathcal{M}_k(\xim)}{\mathcal{M}_k(0)},
\qquad
\widetilde{\QFI}_{\chi\chi}(\xim)
=
\frac{\QFI_{\chi\chi}^{\rm mode}(\xim)}
{\QFI_{\chi\chi}^{\rm mode}(0)}
=
\etam^2,
\label{eq:normalized_resources}
\end{equation}
where $\mathcal{M}_k\in\{\EOF,\QD,\LQU\}$. We do not normalize the dephasing-rate QFI by its value at $\xim=0$, because $\QFI_{\Gamma_{\mathrm m}\Gamma_{\mathrm m}}^{\rm mode}$ has the ideal boundary divergence displayed in Eq.~\eqref{eq:qfi_gamma_mode}.
A useful numerical diagnostic is the Pearson coefficient
\begin{equation}
r_{k,\chi\chi}
=
\frac{
\sum_j
\left(\widetilde{\mathcal{M}}_{k,j}
-\overline{\widetilde{\mathcal{M}}}_k\right)
\left(\widetilde{\QFI}_{\chi\chi,j}
-\overline{\widetilde{\QFI}}_\chi\right)
}{
\sqrt{\sum_j
\left(\widetilde{\mathcal{M}}_{k,j}
-\overline{\widetilde{\mathcal{M}}}_k\right)^2}
\sqrt{\sum_j
\left(\widetilde{\QFI}_{\chi\chi,j}
-\overline{\widetilde{\QFI}}_\chi\right)^2}
}.
\label{eq:resource_qfi_corr}
\end{equation}
A large value of $r_{k,\chi\chi}$ indicates that a resource and the phase QFI degrade similarly over the sampled channel, but it does not establish a universal metrological equivalence. The coefficient depends on the selected $\xim$ interval, grid, and weighting and can be large simply because all curves are monotonic. The comparison is specific to the coherent-phase encoding quantified by Eq.~\eqref{eq:resource_qfi_identities}; the sampling protocol should be stated in the caption or data archive.
\begin{figure*}
		{{\begin{minipage}[t]{.33\linewidth}
					\centering
					(a)\par\vspace{1mm}
\includegraphics[width=\linewidth]{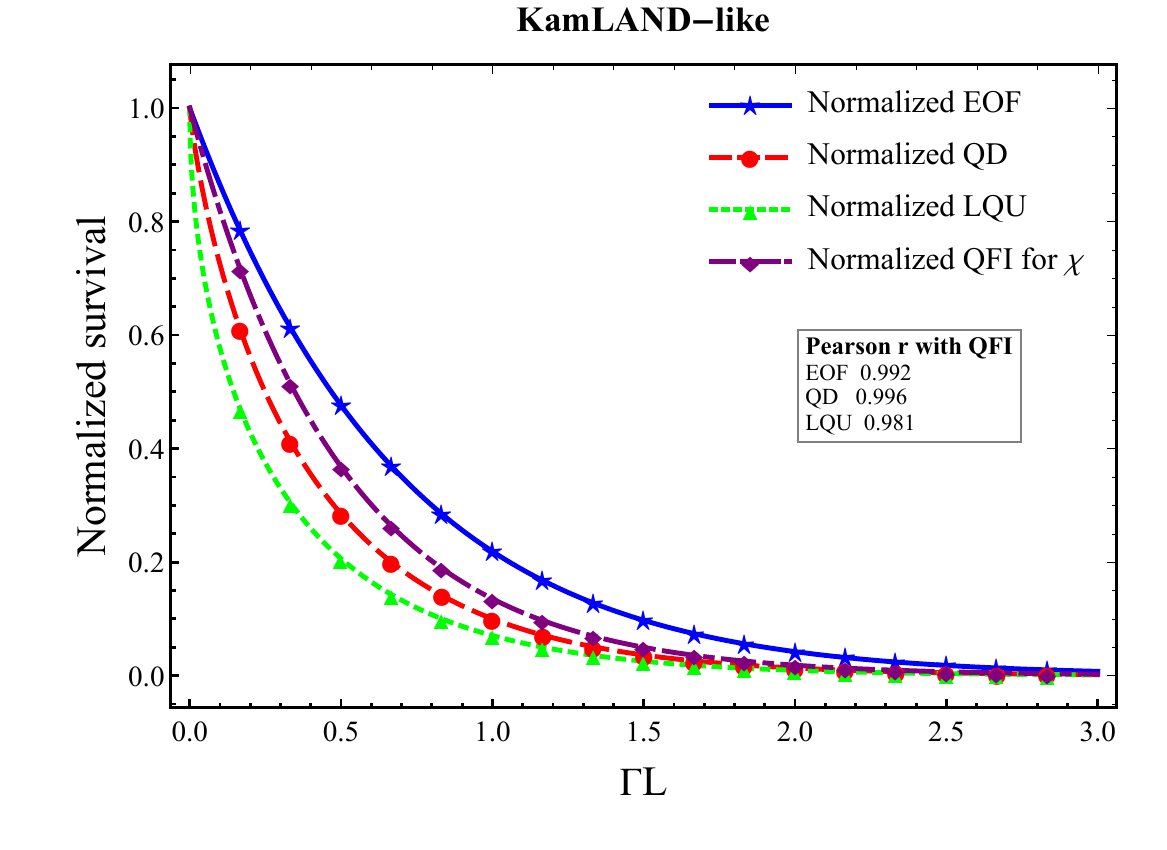}
				\end{minipage}\hfill
				\begin{minipage}[t]{.33\linewidth}
					\centering
					(b)\par\vspace{1mm}
\includegraphics[width=\linewidth]{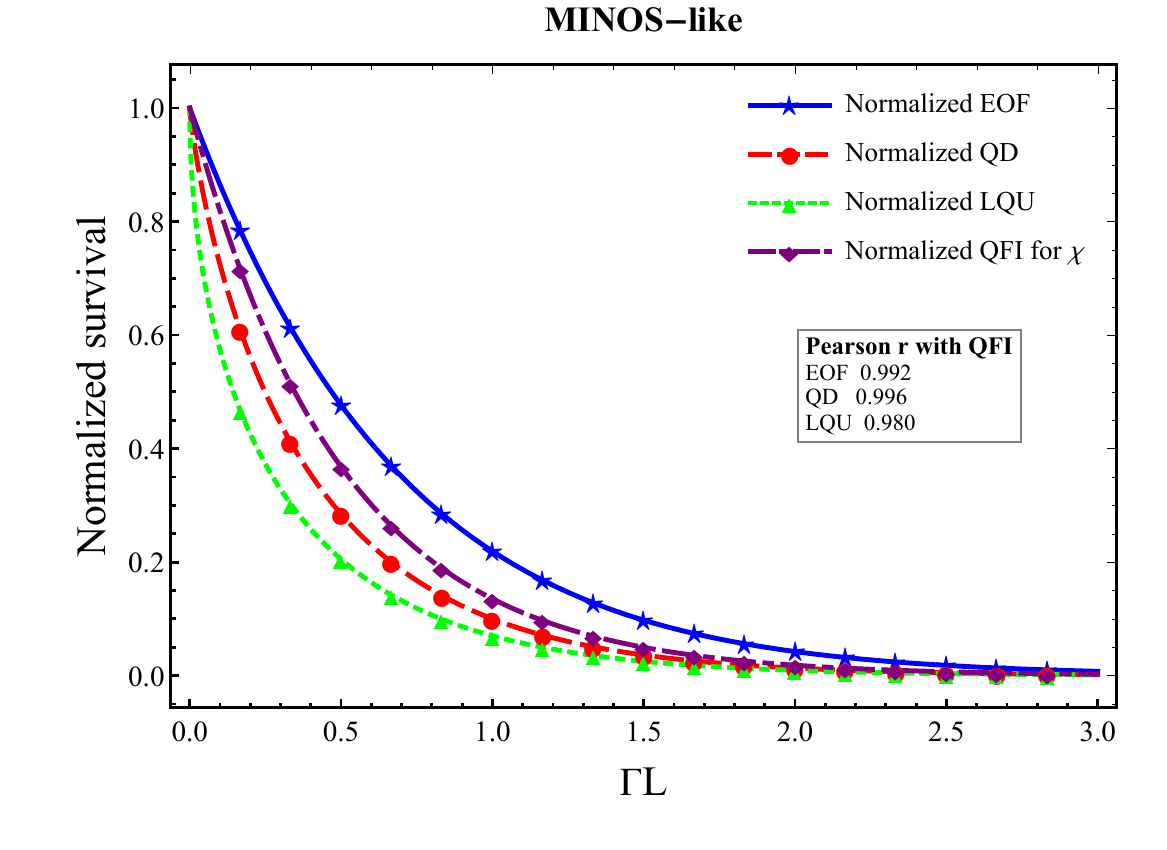}
		\end{minipage}}
	\begin{minipage}[t]{.33\linewidth}
		\centering
		(c)\par\vspace{1mm}
\includegraphics[width=\linewidth]{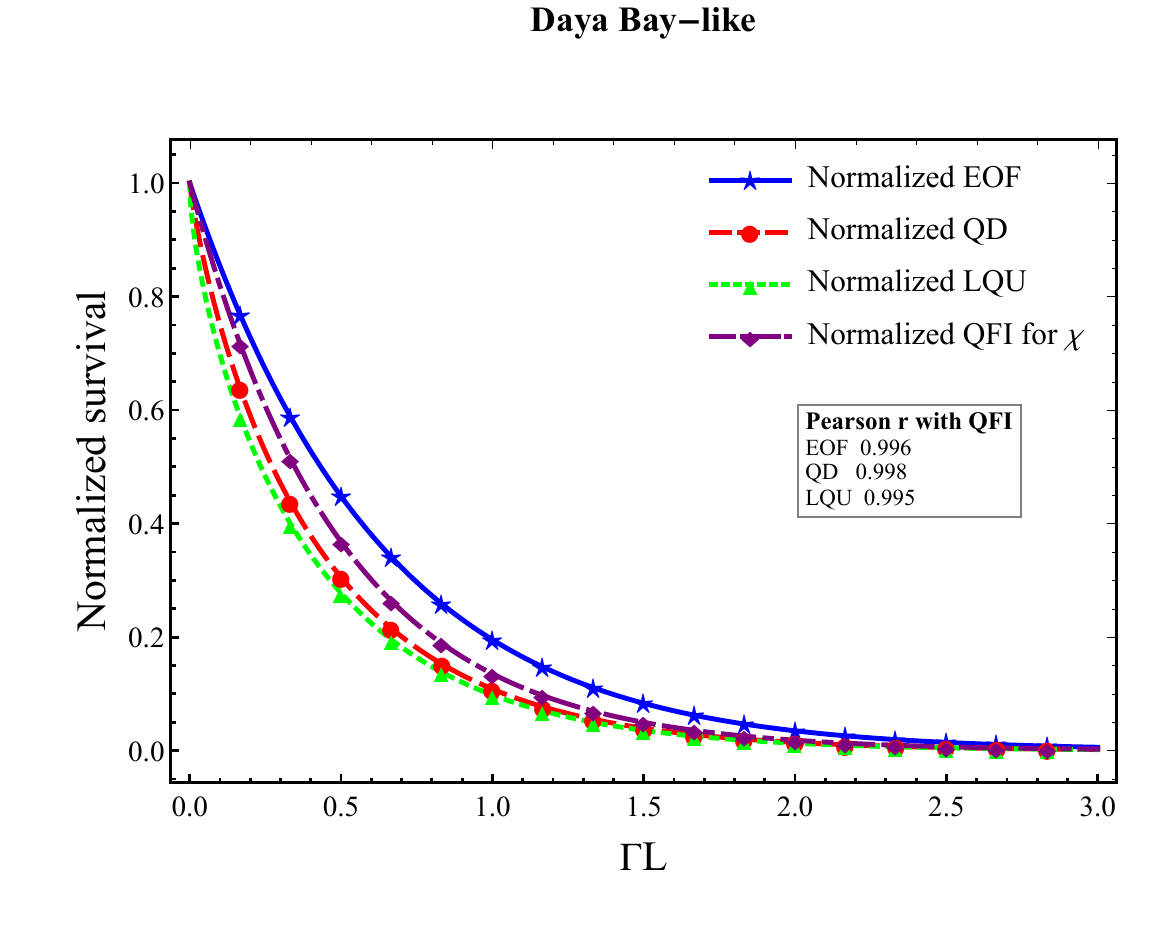}
\end{minipage}}
		\caption{Normalized survival of the entanglement of formation, projective quantum discord, and local quantum uncertainty, compared with the normalized mode-state QFI for the coherent phase $\chi$, $\widetilde{\QFI}_{\chi\chi}=\etam^2$, as functions of the dimensionless mode-dephasing strength $\xim=\Gamma L=-\ln\etam$, for the (a) KamLAND-like, (b) MINOS-like, and (c) Daya Bay-like regimes of Table~\ref{tab:benchmarks}.}
		\label{fig:resource_qfi_survival}
	\end{figure*}
Figure~\ref{fig:resource_qfi_survival} compares, for each benchmark regime, the normalized survival of EOF, QD, and LQU with the normalized phase QFI $\widetilde{\QFI}_{\chi\chi}$. The curves decay on comparable but not identical scales, which is quantified by the Pearson coefficient in Eq.~\eqref{eq:resource_qfi_corr}. This comparison concerns coherent-phase information and should not be interpreted as a normalization of the divergent noise-parameter QFI at $\Gamma=0$.
\subsection{Two-flavor benchmark inputs and scans}

The representative two-flavor regimes used in the analysis are summarized in Table~\ref{tab:benchmarks} and are loosely modeled on the KamLAND~\cite{KamLAND2003}, MINOS~\cite{MINOS2006}, and Daya Bay~\cite{DayaBay2012,DayaBay2018} disappearance measurements.  They should be interpreted as effective channels rather than as complete experimental simulations.

\begin{table*}

\caption{Effective two-flavor inputs used in the analytical scans. The nominal $L/E$ values identify reference points; the regimes are controlled benchmarks rather than complete experimental simulations.}
\label{tab:benchmarks}
\centering
\scriptsize
\setlength{\tabcolsep}{3pt}
\begin{tabular}{@{}p{0.15\linewidth}p{0.24\linewidth}cccc@{}}
\toprule
Regime & Channel & Effective angle & Splitting (eV$^2$) & $(L/E)_{\rm nom}$ & Scan range\\
 & & & & (km/GeV) & (km/GeV)\\
\midrule
KamLAND-like & $\bar\nu_e$ disappearance & $33.44^\circ$ & $7.42\times10^{-5}$ & $6.00\times10^4$ & $0$--$1.00\times10^5$\\
MINOS-like & $\nu_\mu$ disappearance & $49.20^\circ$ & $2.50\times10^{-3}$ & $245$ & $0$--$400$\\
Daya Bay-like & $\bar\nu_e$ disappearance & $8.57^\circ$ & $2.50\times10^{-3}$ & $412.5$ & $0$--$600$\\
\bottomrule
\end{tabular}
\end{table*}
The three regimes differ mostly through the accessible amplitude $A=\sin^2(2\theta)$.  A MINOS-like atmospheric channel is close to maximal and can generate nearly maximal mode correlations.  A KamLAND-like solar channel generates intermediate correlations.  A Daya Bay-like reactor channel is limited by the small value of $\theta_{13}$, even though it is highly precise experimentally because of high statistics and controlled systematics.
    
\begin{figure*}
			{{\begin{minipage}[t]{.32\linewidth}
						\centering
					(a)\par\vspace{1mm}
\includegraphics[width=\linewidth]{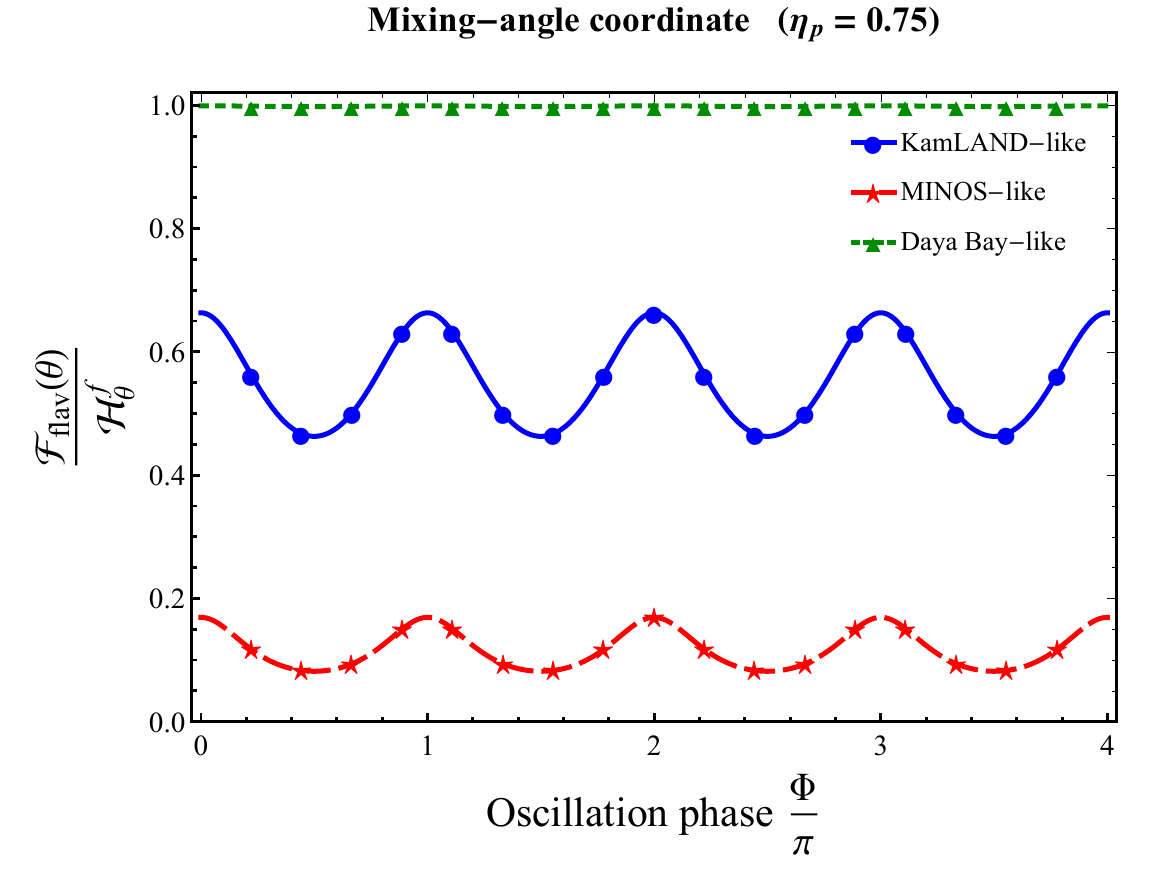}
					\end{minipage}\hfill
					\begin{minipage}[t]{.32\linewidth}
						\centering
					(b)\par\vspace{1mm}
\includegraphics[width=\linewidth]{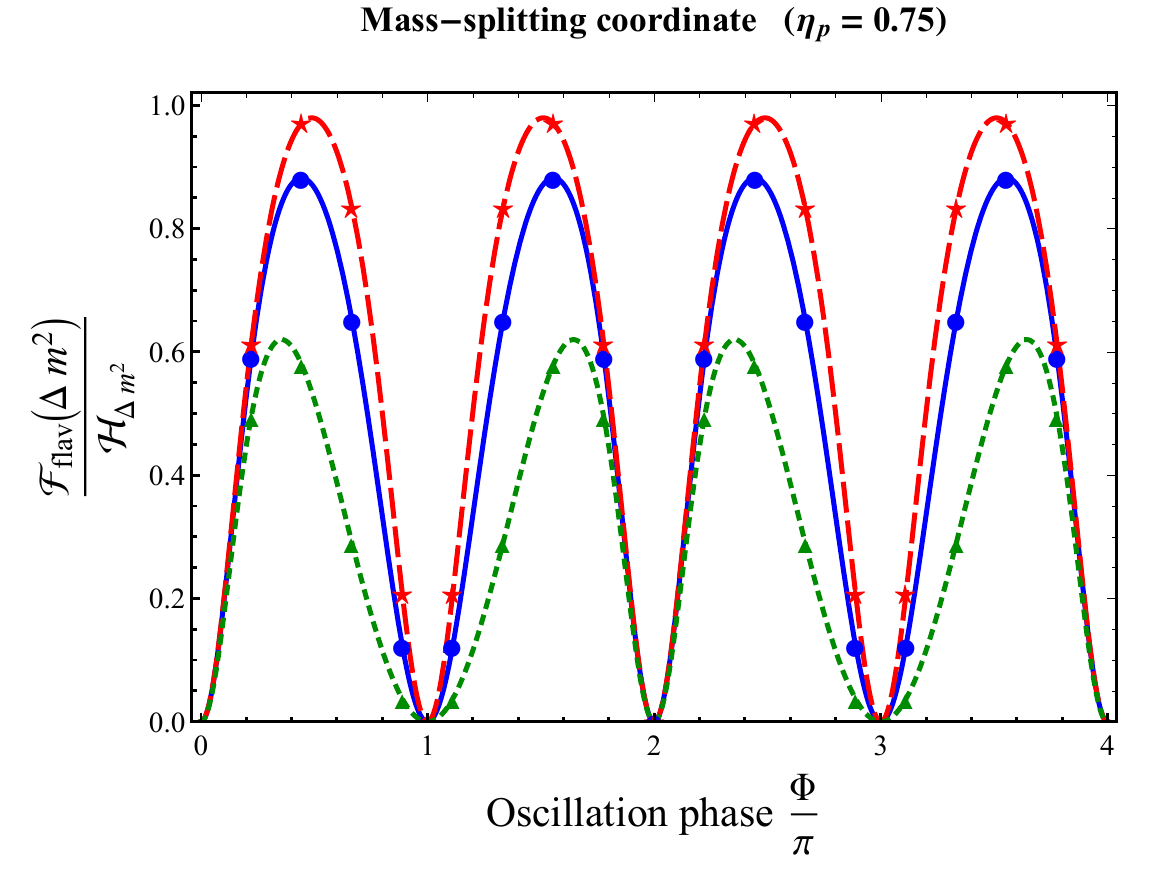}
			\end{minipage}
            \begin{minipage}[t]{.32\linewidth}
						\centering
					(c)\par\vspace{1mm}
\includegraphics[width=\linewidth]{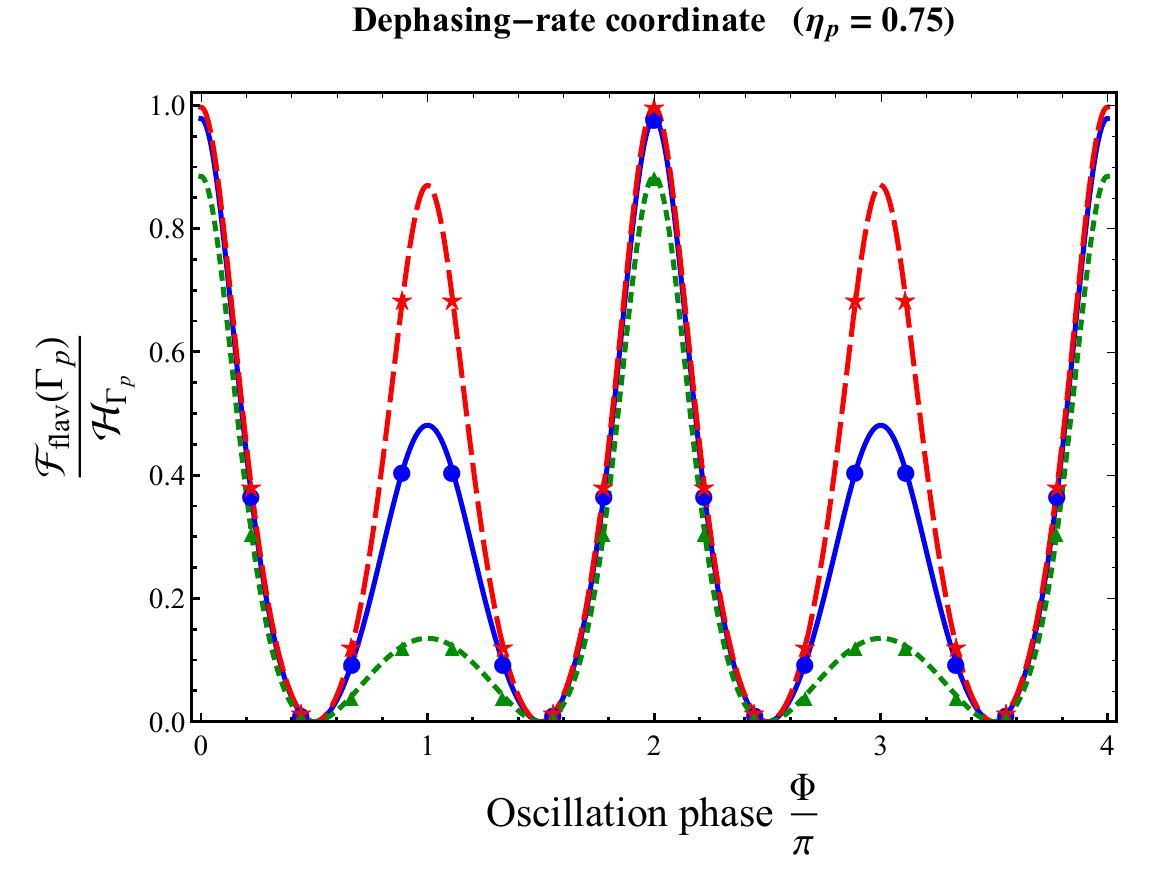}
			\end{minipage}
            }}
\caption{Dimensionless flavor accessibility as a function of the oscillation phase $\Phi/\pi$, evaluated at fixed $\eta_{\rm p}=0.75$ for the KamLAND-like, MINOS-like, and Daya Bay-like benchmark configurations. Panels show the accessibility ratios for the (a) mixing-angle coordinate $\theta$, (b) mass-squared splitting coordinate $\Delta m^2$, and (c) dephasing-rate coordinate $\Gamma_{\rm p}$. The horizontal axis is phase rather than an experiment-specific $L/E$ scan.} 
\label{fig:flavor_accessibility}
\end{figure*}
Figure~\ref{fig:flavor_accessibility} summarizes the dimensionless flavor-accessibility ratios for the three benchmark regimes of Table~\ref{tab:benchmarks} at fixed $\eta_{\rm p}=0.75$.\\
Panel~(a) shows that flavor accessibility for the mixing-angle coordinate depends strongly on the effective mixing amplitude. The Daya Bay-like regime remains close to the basis-matched quantum limit over the displayed phase range, whereas the KamLAND-like regime shows intermediate accessibility. By contrast, the near-maximal MINOS-like configuration exhibits a strongly reduced ratio $\mathcal{F}_{\mathrm{flav}}(\theta)/\mathcal{H}^{(f)}_{\theta}$. This behavior follows from the factor $1-A=\cos^{2}(2\theta)$ entering Eq.~(82): near maximal mixing, the flavor probability becomes locally weakly sensitive to variations of the mixing angle even though state-level information remains available.
\\
Panel~(b) presents the corresponding accessibility for estimating the mass-squared splitting $\Delta m^2$, based on Eqs.~\eqref{eq:qfi-dm2-dephasing}, \eqref{eq:fi-dm2-dephasing}, and \eqref{eq:efficiency-dm2-dephasing}. Since the ratio $\mathcal{F}_{\rm flav}(\Delta m^2)/\mathcal{H}_{\Delta m^2}$ contains a $\sin^2\Phi$ dependence, the flavor accessibility vanishes at the relevant oscillation extrema for all three benchmark regimes. At these phase points, the local derivative of the transition probability with respect to $\Delta m^2$ vanishes, so flavor counting carries no local information on the mass-splitting parameter.
\\
Panel~(c) displays the accessibility associated with the dephasing-rate parameter $\Gamma_{\rm p}$, as described by Eq.~\eqref{eq:efficiency-gamma-dephasing}. In this case, $\mathcal{F}_{\rm flav}(\Gamma_{\rm p})$ is proportional to $\cos^2\Phi$, whereas the corresponding quantum Fisher information does not vanish at $\Phi=\pi/2$. Consequently, flavor measurements become locally insensitive to the dephasing rate at specific phase points, even though those regions may retain substantial quantum information. Taken together, panels~(a)--(c) therefore highlight the parameter-dependent character of flavor accessibility and provide a direct illustration of the operational trade-off discussed in Sec.~\ref{sec:discussion}.

\subsection{Correlation and metrology workflow}

For each benchmark, the workflow is:
\begin{enumerate}
\item Choose $(\theta,\dmsq)$ and compute $\Phi=\dmsq L/(2E)$.
\item Compute $P=\sin^2(2\theta)\sin^2(\Phi/2)$ and $S=1-P$.
\item Construct $\rho_{\etam}$ from Eq.~\eqref{eq:rho_eta} for a chosen mode-coherence profile $\etam(L,E)$. This profile is not equated to $\etap$ unless a microscopic channel relation is specified.
\item Evaluate EOF from Eq.~\eqref{eq:eof_eta}, discord from Eq.~\eqref{eq:discord_eta}, and LQU from Eq.~\eqref{eq:lqu_eta}.
\item Evaluate QFI from Eq.~\eqref{eq:pure_qfi}, Eq.~\eqref{eq:spectral_qfi}, or Eq.~\eqref{eq:qubit_qfi}, depending on whether the state is pure or dephased.
\item Evaluate flavor FI from the relevant flavor probabilities and compare $\CFI_{\rm flav}/\QFI$.
\end{enumerate}
This protocol deliberately separates state preparation, open-system propagation, and measurement.  That separation is essential for interpreting a precision limit: a small FI may arise because the state carries little information, because the measurement is not optimal, or because decoherence has removed the relevant coherence.

\begin{figure}[!htbp]
\centering
\includegraphics[width=0.8\linewidth]{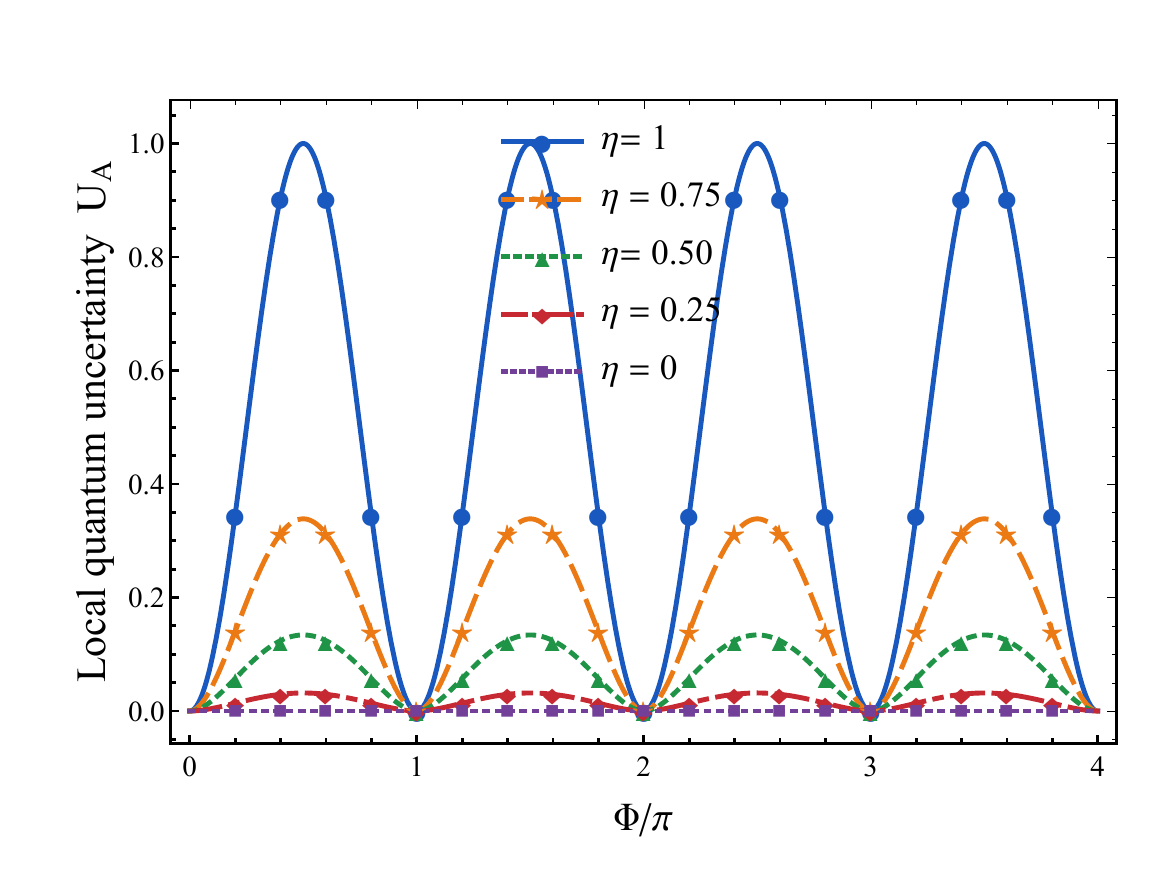}
\caption{Auxiliary two-flavor diagnostic: local quantum uncertainty for a maximal-mixing channel as a function of oscillation phase and mode-dephasing amplitude $\etam$.  LQU tracks the coherent part of the transition-superposition state and vanishes when $\etam\to0$ or when the state is localized in one flavor mode.}
\label{fig:lqu_dephasing}
\end{figure}

\subsection{Local sensitivity derivatives}

The local response of a resource $\mathcal{M}(P,\etam)$ to a physical parameter $x$ is
\begin{equation}
    \frac{\partial \mathcal{M}}{\partial x}=\frac{\partial\mathcal{M}}{\partial P}\frac{\partial P}{\partial x}
    +\frac{\partial\mathcal{M}}{\partial\eta}\frac{\partial\eta}{\partial x}.
    \label{eq:chain_resource}
\end{equation}
For the unitary probability $P=A\sin^2(\Phi/2)$,
\begin{align}
    \partial_\theta P&=2\sin(4\theta)\sin^2\frac{\Phi}{2},
    \\
    \partial_{\dmsq}P&=\frac{A L}{4E}\sin\Phi,
    \\
    \partial_L P&=\frac{A\dmsq}{4E}\sin\Phi,
    \\
    \partial_E P&=-\frac{A\dmsq L}{4E^2}\sin\Phi.
\end{align}
For a phenomenological mode profile $\etam=\exp[-\Gamma_{\mathrm m,0}(E/E_0)^nL]$,
\begin{align}
    \partial_L\etam&=-\Gamma_{\mathrm m}(E)\etam,\\
    \partial_{\Gamma_{\mathrm m,0}}\etam&=-(E/E_0)^nL\etam,\\
    \partial_E\etam&=-\etam L\Gamma_{\mathrm m}(E)\frac{n}{E}.
\end{align}
These derivatives are the basis for comparing ordinary probability sensitivity with the sensitivity of correlation measures.  In the mode-dephasing model, probability derivatives do not contain $\partial_x\etam$ unless the physical dephasing is modeled in the oscillation basis; by contrast, EOF, QD, and LQU respond directly to $\etam$.

\FloatBarrier
\section{Discussion}
\label{sec:discussion}

\begin{figure*}[htbp]
\centering
\begin{tikzpicture}[
    font=\small,
    >=Latex,
    node distance=7mm and 8mm,
    mainbox/.style={
        draw=NJPBlue,
        rounded corners=3pt,
        line width=0.9pt,
        fill=NJPBlueLight,
        align=center,
        inner sep=6pt,
        text width=0.80\textwidth
    },
    diagbox/.style={
        draw=NJPBlue,
        rounded corners=3pt,
        line width=0.8pt,
        fill=white,
        align=center,
        inner sep=6pt,
        text width=0.24\textwidth,
        minimum height=1.95cm
    },
    finalbox/.style={
        draw=NJPBlue,
        rounded corners=3pt,
        line width=0.9pt,
        fill=NJPGray,
        align=center,
        inner sep=6pt,
        text width=0.80\textwidth
    },
    flow/.style={
        -{Latex[length=2.4mm,width=1.6mm]},
        line width=0.9pt,
        draw=NJPBlue
    },
    auxflow/.style={
        -{Latex[length=2.2mm,width=1.5mm]},
        line width=0.8pt,
        draw=NJPBlue,
        dashed
    }
]


\node[mainbox] (prep) {
    \textbf{\textcolor{NJPBlue}{State preparation and parameter encoding}}\\
    Eqs.~\eqref{eq:three_state} and \eqref{eq:mode_state}
};

\node[mainbox, below=of prep] (prop) {
    \textbf{\textcolor{NJPBlue}{Open-system propagation}}\\
    Propagation-basis and flavor-mode dephasing\\
    Eqs.~\eqref{eq:mass_dephasing} and \eqref{eq:rho_eta}
};


\coordinate (fork) at ($(prop.south)+(0,-7mm)$);


\node[
    diagbox,
    anchor=north,
    at={($(fork)+(-0.29\textwidth,-8mm)$)}
] (qfi) {
    \textbf{\textcolor{NJPBlue}{Intrinsic state information}}\\
    QFI / QFIM\\
    Eqs.~\eqref{eq:qcrb}, \eqref{eq:mode_qfim}
};

\node[
    diagbox,
    anchor=north,
    at={($(fork)+(0,-8mm)$)}
] (res) {
    \textbf{\textcolor{NJPBlue}{Quantum-resource survival}}\\
    EOF, QD, LQU\\
    Eqs.~\eqref{eq:eof_eta},
    \eqref{eq:discord_eta},
    \eqref{eq:lqu_eta}
};

\node[
    diagbox,
    anchor=north,
    at={($(fork)+(0.29\textwidth,-8mm)$)}
] (fi) {
    \textbf{\textcolor{NJPBlue}{Accessible flavor information}}\\
    Flavor FI / FIM\\
    Eqs.~\eqref{eq:fi_two_outcome},
    \eqref{eq:threeflavor_flavor_fim}
};


\node[
    finalbox,
    below=18mm of res
] (compat) {
    \textbf{\textcolor{NJPBlue}{Joint-estimation attainability}}\\
    Normalized SLD incompatibility\\
    Eq.~\eqref{eq:sld_compatibility}
};


\coordinate (merge) at ($(compat.north)+(0,7mm)$);


\draw[flow] (prep.south) -- (prop.north);
\draw[flow] (prop.south) -- (fork);


\draw[flow] (fork) -| (qfi.north);
\draw[flow] (fork) -- (res.north);
\draw[flow] (fork) -| (fi.north);


\draw[flow] (qfi.south) |- (merge);
\draw[auxflow] (res.south) -- (merge);
\draw[flow] (fi.south) |- (merge);

\draw[flow] (merge) -- (compat.north);

\end{tikzpicture}

\caption{
Operational map of the manuscript.
After open-system propagation, the analysis splits into three complementary diagnostics:
intrinsic state sensitivity (QFI/QFIM), measurement-accessible flavor information,
and survival of nonclassical resources.
Their synthesis clarifies the origin of precision loss and the attainability of
multiparameter quantum bounds.
}
\label{fig:information_loss_framework}
\end{figure*}
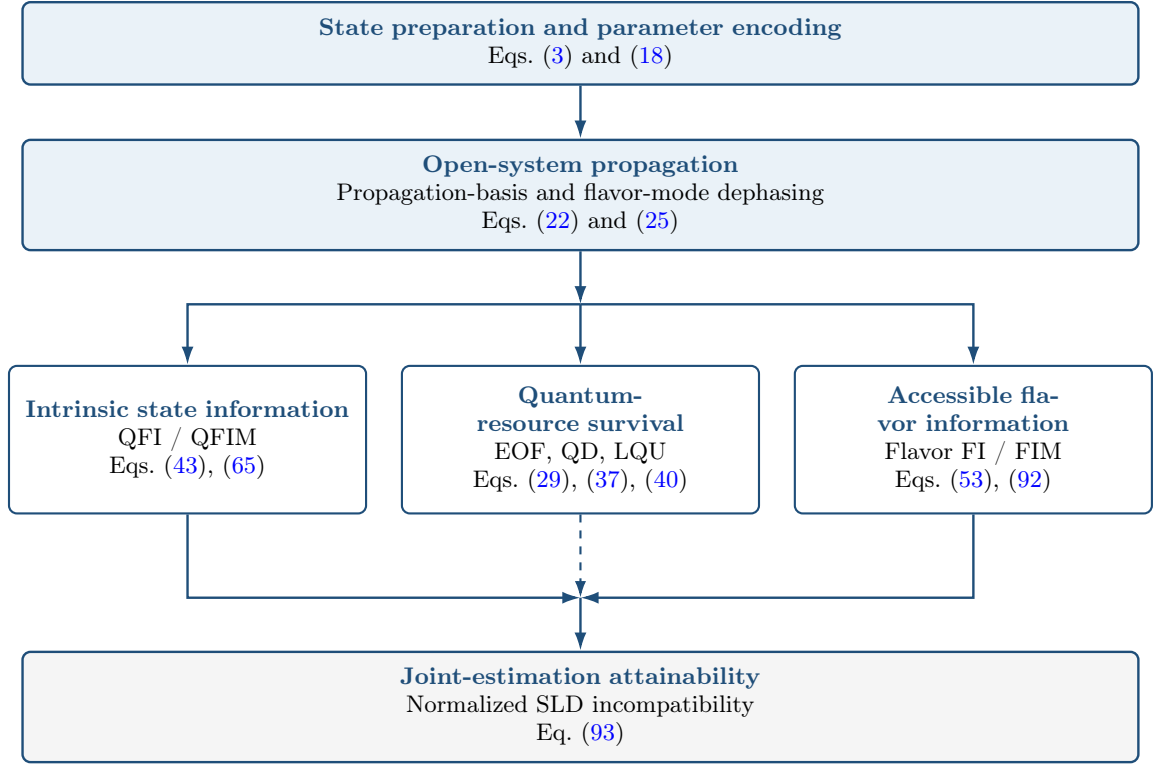
\subsection{Four layers of information loss}

The results establish that four conceptually different quantities must be kept
separate in a decohering oscillation problem. The first is the intrinsic
sensitivity of the propagated state, quantified by the QFI or QFIM. The second
is the information extracted by the chosen flavor measurement, quantified by
the corresponding FI or FIM. The third is the survival of nonclassical
mode resources, represented here by concurrence, EOF, one-sided projective QD,
and LQU. The fourth is the compatibility of the optimal measurements required
for simultaneous estimation. None of these layers can in general be inferred
from another one.

The distinction is already exact in the two-flavor mode model. A population
measurement saturates the $P$ component of the mode-state QFIM, but it contains
no information on either the coherent phase $\chi$ or the mode-dephasing
amplitude $\etam$. Conversely, the off-diagonal sector carries finite phase and
noise information even though the diagonal populations are unchanged. This
provides a direct example in which an experimentally natural measurement is
optimal for one coordinate of a statistical model and completely blind to two
others.

The propagation-basis calculation exposes a complementary limitation. The
flavor FI for the dephasing rate contains a factor $\cos^2\Phi$ and therefore
vanishes at $\Phi=(2k+1)\pi/2$, whereas the propagation-state QFI remains
nonzero. At those phases the state still changes with the damping parameter,
but the diagonal flavor probabilities do not change to first order. The lost
information is therefore measurement-induced rather than absent from the
state. By contrast, the mass-splitting FI contains $\sin^2\Phi$ and vanishes at
probability extrema, where the local event-rate slope is zero. These analytic
zeros demonstrate that the best operating phase depends on the parameter being
estimated; there is no single $L/E$ point that simultaneously optimizes mixing,
phase, splitting, and noise estimation.

\subsection{Resource measures as encoding-specific metrological indicators}

The exact diagonal QFIM of the flavor-mode state gives a precise interpretation
to the resource diagnostics. The identity
$\QFI_{\chi\chi}^{\rm mode}=C_{\etam}^{2}$ shows that concurrence is exactly the
square root of the coherent-phase information for this family. Since EOF is a
monotonic function of concurrence, it is also a monotonic indicator of the same
phase encoding. LQU tracks that information through a mixedness-dependent
factor, and the numerical survival curves confirm a close, but not identical,
decay. The reported Pearson coefficients near unity therefore quantify a model- and sampling-specific co-variation with coherent-phase QFI; they do not establish a universal equality between quantum correlations and metrological usefulness. Their values should be accompanied by the sampled interval, grid, and weighting.

This qualification is essential. A resource may survive while the selected
measurement is unable to access it, and a state may retain information on a
noise parameter even after an entanglement measure has become small. Resource
survival should therefore be interpreted as a structural diagnostic of the
state, not as a substitute for a parameter-specific QFI or for a complete
measurement model.

\subsection{Meaning of the three-flavor benchmarks}

The three-flavor results extend the operational separation to the parameter set
$(\theta_{23},\delta_{\rm CP},\xi_{31},g)$. At
$\Gamma_0=10^{-23}\,\mathrm{GeV}$, the diagonal fixed-coordinate flavor
accessibility for $\delta_{\rm CP}$ is approximately $0.010$ at the DUNE-like
point, $0.013$ at the T2HK-like point, and $0.105$ at the ESSnuSB-inspired
second-maximum vacuum point. Within the monochromatic benchmark, the
second-maximum configuration therefore converts a larger fraction of the
state-level CP information into flavor-counting information. This comparison
should not be read as an experimental ranking: the three points use different
baselines, energies, and matter prescriptions, and they omit fluxes, spectral
weights, detector response, and nuisance parameters.

The same table shows that flavor access is strongly parameter dependent. For
$\theta_{23}$ the diagonal ratios are about $0.30$, $0.66$, and $0.65$ for the
DUNE-like, T2HK-like, and ESSnuSB-inspired points, respectively. For
$\xi_{31}$ they are $0.047$, $0.081$, and $0.095$, while for the logarithmic
decoherence coordinate $g$ they are $0.61$, $0.22$, and $0.21$. The relatively
large DUNE-like value for $g$ is consistent with the stronger accumulated
damping at the longer baseline, whereas the CP coordinate remains poorly
accessed by a single flavor projection at the chosen energy.\\
The multiparameter analysis adds information that diagonal ratios cannot provide. Nevertheless, the single-energy flavor FIM has rank at most two, so its four diagonal entries do not amount to a jointly identifiable four-parameter measurement model. Energy binning and additional channels are required before a flavor-FIM covariance matrix can be compared with a four-parameter quantum bound.\\
The normalized SLD incompatibility between $\theta_{23}$ and
$\xi_{31}$ is approximately $0.93$, $0.98$, and $0.97$ in the three
benchmarks. These values are large even when the corresponding real QFIM
correlations are modest. Statistical correlation and measurement
incompatibility are therefore distinct geometric properties. The inverse QFIM
must consequently be described as an SLD matrix lower bound, not as a set of
simultaneously attainable error bars. A Holevo-bound calculation or an
explicit optimization over compatible measurements is required before making
an operational claim about the best joint precision.

\subsection{Role of baseline and damping}

The known-rate SLD-QCRB penalty increases most strongly at the selected DUNE-like point for large $\Gamma_0$. The larger penalty is consistent with the accumulated factor $\Gamma_0L$, but the benchmark comparison simultaneously changes baseline, energy, oscillation phase, and matter effects. It therefore cannot isolate baseline as the sole cause. A longer baseline may strengthen matter and phase encoding while also increasing exposure to decoherence; the useful regime is determined by the competition among encoding strength, coherence survival, measurement accessibility, and nuisance-parameter correlations.

\subsection{Scope, reproducibility, and required extensions}

The analysis is intentionally state-level. It uses single-energy benchmark
points, constant-density propagation where specified, a common off-diagonal
damping rate, ideal flavor projection, and fixed oscillation inputs. It does
not include fluxes, cross sections, energy migration, efficiencies,
backgrounds, systematic pulls, external priors, or combined neutrino and
antineutrino samples. The ESSnuSB-inspired point is explicitly a vacuum
second-maximum benchmark and should not be interpreted as a complete simulation
of the proposed facility.

Five extensions are particularly important. First, evaluate the full four-parameter QFIM with $g$ treated as an unknown nuisance parameter and report the marginalized standard-parameter penalties. Second, convert the large SLD incompatibility into an attainable multiparameter bound, preferably the Holevo bound or an explicitly optimized compatible measurement family~\cite{Albarelli2019,Xia2023}. Third, replace the common-rate ansatz by independent rates $(\Gamma_{21},\Gamma_{31},\Gamma_{32})$ constrained by complete positivity and test several energy exponents. Fourth, compare a co-rotating matter-eigenstate noise basis with a fixed reference coupling basis. Fifth, perform an event-level extension with energy binning, neutrino and antineutrino samples, realistic matter profiles, fluxes, cross sections, detector response, backgrounds, priors, and systematic nuisance parameters.

The numerical validation diagnostics reported in
Table~\ref{tab:numerical_validation} provide explicit checks of
normalization, Hermiticity, state positivity, QFIM
positive-semidefiniteness, consistency of
$\mathcal{H}-\mathcal{F}^{\rm flav}$, finite-difference stability, and
the expected rank deficiency of the monochromatic flavor FIM at the
three benchmark points. These tests strengthen the reproducibility of
the state-level analysis by showing that the reported
information-geometric quantities are numerically stable within the
adopted precision. A complete reproducibility archive should
additionally provide the source code, input parameter tables, full
numerical matrices, singular values of the flavor FIMs, and
machine-readable data associated with each figure and table.

\FloatBarrier

\FloatBarrier
\section{Conclusions}
\label{sec:conclusion}

This work develops a basis-explicit diagnostic framework for decohering neutrino oscillations. Its first requirement is to keep two statistical models separate. Propagation-basis dephasing suppresses interference before flavor projection and changes the flavor probabilities; effective flavor-mode dephasing suppresses the coherence of a single-particle, two-mode state while keeping its mode populations fixed. The two survival factors therefore should not be identified without a microscopic map.

For the flavor-mode state, the QFIM in $(P,\chi,\etam)$ is diagonal in the regular interior. The identities $\QFI^{\rm mode}_{\chi\chi}=C_{\etam}^{2}$ and $\QFI^{\rm mode}_{\xim\xim}=C_{\etam}^{2}/(1-\etam^{2})$ give the resource measures a precise, encoding-dependent interpretation: EOF is a monotonic function of coherent-phase information, while LQU is a mixedness-dependent fraction of it. The divergence at vanishing dephasing is a nonregular support-changing boundary and should not be interpreted through the ordinary local Cram\'er--Rao bound without a finite-resolution statistical model.

For propagation-basis dephasing, a valid QFI--FI comparison requires one explicitly defined preparation, propagation, noise-basis, and measurement model. When flavor projectors are treated as fixed laboratory outcomes and all mixing-angle dependence is assigned to the state map, the flavor FI and basis-matched flavor-basis QFI both equal $16$ at the coherent first maximum for nonmaximal mixing. The accessibility ratios also show that mass-splitting information is recovered on oscillation slopes, whereas flavor sensitivity to the dephasing rate vanishes at $\cos\Phi=0$ even though the state-level QFI can remain finite.

The monochromatic three-flavor benchmarks extend this separation to $(\theta_{23},\delta_{\rm CP},\xi_{31},g)$. At $\Gamma_0=10^{-23}\,\mathrm{GeV}$, the fixed-coordinate CP accessibility ratios are $0.010$, $0.013$, and $0.105$ for the DUNE-like, T2HK-like, and ESSnuSB-inspired vacuum points, while the dominant $\theta_{23}$--$\xi_{31}$ SLD incompatibilities are $0.93$, $0.98$, and $0.97$. These values demonstrate that state sensitivity, flavor accessibility, resource survival, and joint attainability are distinct properties. They do not rank the experiments or constitute event-level forecasts. In addition, the single-energy flavor FIM has rank at most two, so four-parameter joint identification requires multiple energies or other independent measurement settings.

The framework is strongest as an analytic and information-geometric benchmark. Accordingly, the present numerical study intentionally preserves one fixed legacy oscillation-input set across all curves and tables, rather than mixing numerical results generated with different generations of global-fit inputs. The absolute benchmark values should
therefore be interpreted conditionally on this reference point. A dedicated robustness study based on a current global-fit ensemble would constitute a separate analysis and would require the consistent regeneration of every dependent numerical result. The numerical validation diagnostics in Table~\ref{tab:numerical_validation}
confirm normalization, Hermiticity, positivity, finite-difference stability, and the expected rank deficiency of the flavor FIM for the
present benchmark. Further extensions include releasing the complete code and machine-readable numerical archive, testing the dependence on the assumed dephasing basis, treating the dephasing rate as a
nuisance parameter, and replacing the SLD matrix bound by an
attainable multiparameter benchmark. An energy-binned
neutrino--antineutrino analysis including realistic detector and
systematic effects would then establish whether the state-level
accessibility gaps identified here persist at the experimental level.
\appendix
\FloatBarrier
\section{Derivation of the diagonal mode-state QFIM}
\label{app:mode_qfim}
For Eq.~\eqref{eq:bloch_eta},
\begin{equation}
1-|\bm r|^2=4P(1-P)(1-\etam^2).
\end{equation}
Direct differentiation gives
\begin{align}
\partial_P\bm r\cdot\partial_\chi\bm r&=0, &
\partial_\chi\bm r\cdot\partial_{\etam}\bm r&=0,\\
\partial_P\bm r\cdot\partial_{\etam}\bm r
+\frac{(\bm r\cdot\partial_P\bm r)
(\bm r\cdot\partial_{\etam}\bm r)}{1-|\bm r|^2}&=0.
\end{align}
The remaining diagonal terms reduce to
\begin{align}
\QFI_{PP}&=\frac{1}{P(1-P)},\\
\QFI_{\chi\chi}&=4\etam^2P(1-P),\\
\QFI_{\etam\etam}&=\frac{4P(1-P)}{1-\etam^2},
\end{align}
which proves Eq.~\eqref{eq:mode_qfim}.

\FloatBarrier
\section{Derivation of the two-flavor amplitudes}

Using Eq.~\eqref{eq:two_rotation}, the initial flavor state is
\begin{equation}
    \ket{\nu_\alpha}=\cos\theta\ket{\nu_i}+\sin\theta\ket{\nu_j}.
\end{equation}
After propagation and removal of the common $\nu_i$ phase,
\begin{equation}
    \ket{\psi(L)}=\cos\theta\ket{\nu_i}+\sin\theta\ee^{-\ii\Phi}\ket{\nu_j}.
\end{equation}
The inverse rotation is
\begin{equation}
    \ket{\nu_i}=\cos\theta\ket{\nu_\alpha}-\sin\theta\ket{\nu_\beta},
    \qquad
    \ket{\nu_j}=\sin\theta\ket{\nu_\alpha}+\cos\theta\ket{\nu_\beta}.
\end{equation}
Substitution gives
\begin{align}
    A_{\alpha\alpha}&=\cos^2\theta+\sin^2\theta\ee^{-\ii\Phi},\\
    A_{\alpha\beta}&=\sin\theta\cos\theta(\ee^{-\ii\Phi}-1).
\end{align}
Therefore
\begin{align}
    \abs{A_{\alpha\beta}}^2&=\sin^2\theta\cos^2\theta\abs{\ee^{-\ii\Phi}-1}^2\\
    &=4\sin^2\theta\cos^2\theta\sin^2\frac{\Phi}{2}\\
    &=\sin^2(2\theta)\sin^2\frac{\Phi}{2}.
\end{align}

\FloatBarrier
\section{Derivation of the LQU formula}

The nonzero block of $\rho_{\etam}$ is
\begin{equation}
    \rho_b=\begin{pmatrix}S&\etam z\\ \etam z^*&P\end{pmatrix},
    \qquad S+P=1,
    \qquad \abs{z}=\sqrt{SP}.
\end{equation}
Its determinant is
\begin{equation}
    d=SP(1-\etam^2).
\end{equation}
For a positive $2\times2$ matrix of unit trace,
\begin{equation}
    \sqrt{\rho_b}=\frac{\rho_b+\sqrt{d}\,I_2}{\sqrt{1+2\sqrt d}}.
\end{equation}
Let the off-diagonal entry of $\sqrt{\rho_b}$ be $b$.  Then
\begin{equation}
    \abs{b}^2=\frac{\etam^2SP}{1+2\sqrt{SP(1-\etam^2)}}.
\end{equation}
In the full two-qubit basis, only $W_{zz}$ is nonzero in Eq.~\eqref{eq:lqu_general}, and
\begin{equation}
    W_{zz}=1-4\abs{b}^2.
\end{equation}
Consequently,
\begin{equation}
    \LQU_A=1-W_{zz}=\frac{4\etam^2SP}{1+2\sqrt{SP(1-\etam^2)}}.
\end{equation}
This proves Eq.~\eqref{eq:lqu_eta}.

\FloatBarrier
\section{Conditional entropy for the discord calculation}

For completeness, we give the explicit ingredients of Eq.~\eqref{eq:discord_cond}.  By a local phase rotation, take $z\geq0$.  Define the projectors on subsystem $B$ by the angle $\vartheta$ and set $u=\cos(\vartheta/2)$, $v=\sin(\vartheta/2)$.  The two outcome probabilities are
\begin{align}
    q_0&=Pu^2+Sv^2,\\
    q_1&{}=Pv^2+Su^2.
\end{align}
The unnormalized conditional states of subsystem $A$ have matrix elements
\begin{align}
    \tilde\rho_{A|0}&=\begin{pmatrix}Sv^2&\etam z uv\\ \etam z uv&Pu^2\end{pmatrix},\\
    \tilde\rho_{A|1}&=\begin{pmatrix}Su^2&-\etam z uv\\ -\etam z uv&Pv^2\end{pmatrix}.
\end{align}
The normalized states are $\rho_{A|k}=\tilde\rho_{A|k}/q_k$.  Their Bloch-vector lengths are
\begin{align}
    \abs{\bm r_0}&=\frac{\sqrt{(Sv^2-Pu^2)^2+4\etam^2SPu^2v^2}}{q_0},\\
    \abs{\bm r_1}&=\frac{\sqrt{(Su^2-Pv^2)^2+4\etam^2SPu^2v^2}}{q_1}.
\end{align}
Substitution into Eq.~\eqref{eq:discord_cond} gives a one-dimensional minimization for the exact projective discord.

\FloatBarrier
\section{Pure two-flavor QFI derivation}

For Eq.~\eqref{eq:two_mass_state},
\begin{align}
    \ket{\partial_\theta\psi}&=-\sin\theta\ket{\nu_i}+\cos\theta\ee^{-\ii\Phi}\ket{\nu_j},\\
    \braket{\psi}{\partial_\theta\psi}&=0,\\
    \braket{\partial_\theta\psi}{\partial_\theta\psi}&=1.
\end{align}
Equation~\eqref{eq:pure_qfi} gives $\QFI(\theta)=4$.  Similarly,
\begin{align}
    \ket{\partial_\Phi\psi}&=-\ii\sin\theta\ee^{-\ii\Phi}\ket{\nu_j},\\
    \braket{\partial_\Phi\psi}{\partial_\Phi\psi}&=\sin^2\theta,\\
    \abs{\braket{\psi}{\partial_\Phi\psi}}^2&=\sin^4\theta,
\end{align}
so that
\begin{equation}
    \QFI(\Phi)=4\sin^2\theta\cos^2\theta=\sin^2(2\theta).
\end{equation}

\FloatBarrier
\section{Notes on units}

In practical oscillation plots one often writes
\begin{equation}
    \Delta=1.267\left(\frac{\dmsq}{\mathrm{eV}^2}\right)
    \left(\frac{L}{\mathrm{km}}\right)
    \left(\frac{\mathrm{GeV}}{E}\right).
\end{equation}
The formulas in the main text use natural units with $\Delta=\dmsq L/(4E)$ and $\Phi=2\Delta$. For the damping exponent, the baseline must also be converted consistently: $1\,\mathrm{km}=5.06773071616\times10^{18}\,\mathrm{GeV}^{-1}$. Hence $\Gamma[\mathrm{GeV}]L[\mathrm{km}]$ is implemented as $\Gamma L\times5.06773071616\times10^{18}$.

\section*{Acknowledgments} 

The research by Berihu Teklu was funded by Khalifa University of Science and Technology through the Project ID: KU-INT-RIG-2024-8474000739 and was supported by KU Research Center for Advanced Intelligent Systems (AIS), Khalifa University of Science and Technology (KU-AIS).

\section*{Data Availability} 

The data that support the findings of this study are available upon reasonable request from the authors.
The data comprises solutions to equations which are stated in the text and thus easily reproducible by
other researchers.

\end{document}